\documentclass[12pt,preprint2]{aastex}

\newcommand{\kms}{\hbox{${\rm km}\:{\rm s}^{-1}\;$}}

\newcommand{\kmso}{\hbox{${\rm km}\:{\rm s}^{-1}$}}
\newcommand{\teff}{$T_{\rm eff}\;$}

\newcommand{\loggl}{$\log (g/{\rm cm~s}^2)$}

\usepackage{longtable}
\usepackage{lscape}

\shorttitle{Signatures of planets in solar analogs}
\shortauthors{Gonz\'alez Hern\'andez et al.}

\begin{document}

\title{Searching for the signatures of terrestial planets in solar
analogs} 

\author{J. I. Gonz\'alez Hern\'andez\altaffilmark{1,2},
G.~Israelian\altaffilmark{1}, N.~C. Santos\altaffilmark{3,4}, S.
Sousa\altaffilmark{3}, E.~Delgado-Mena\altaffilmark{1}, 
V.~Neves\altaffilmark{3}, \& S. Udry\altaffilmark{5}} 
 
\email{jonay@iac.es}

\altaffiltext{1}{Instituto de Astrof{\'\i}sica de Canarias, C/ Via
L\'actea s/n, 38200 La Laguna, Spain: jonay@iac.es}
\altaffiltext{2}{Dpto. de Astrof\'{\i}sica y Ciencias de la
Atm\'osfera, Facultad de Ciencias F\'{\i}sicas, Universidad
Complutense de Madrid, E-28040 Madrid, Spain}
\altaffiltext{3}{Centro de Astrof\'isica, Universidade do Porto, Rua
das Estrelas, 4150-762 Porto, Portugal} 
\altaffiltext{4}{Departamento de F\'{\i}sica e Astronomia, Faculdade 
de Ci\^encias, Universidade do Porto, Portugal} 
\altaffiltext{5}{Observatoire Astronomique de l'Universit\'e de
Gen\`eve, 51 Ch. des Maillettes, -Sauverny- Ch1290, Versoix, 
Switzerland} 

\begin{abstract}

We present a fully differential chemical abundance analysis using 
very high-resolution ($\lambda/\delta\lambda \gtrsim 85,000$) and 
very high signal-to-noise (S/N~$\sim 800$ on average) HARPS and 
UVES spectra of 7 solar twins and 95 solar analogs,    
24 are planet hosts and 71 are stars without detected planets. 
The whole sample of solar analogs provide very accurate Galactic 
chemical evolution trends in the metalliciy range 
$-0.3<{\rm [Fe/H]}<0.5$.
Solar twins with and without planets show similar mean abundance 
ratios. We have also analysed a sub-sample of 28 solar analogs,
14 planet hosts and 14 stars without known planets, 
with spectra at S/N~$\sim 850$ on average, 
in the metallicity range $0.14<{\rm [Fe/H]}<0.36$ and 
find the same abundance pattern for both samples of stars with and 
without planets. This result does not depend on either the planet mass, from 
7 Earth masses to 17.4 Jupiter masses,
or the orbital period of the planets, from 3 to 4300 days. 
In addition, we have derived the slope of the abundance ratios as 
a function of the condensation temperature for each star and again
find similar distributions of the slopes for both stars 
with and without planets. In particular, the peaks of these two 
distributions are placed at a similar value but with opposite sign as
that expected from a possible signature of terrestial planets. 
In particular, two of the planetary systems in this sample, containing
each of them a Super-Earth like planet, show slope values very close 
to these peaks which may suggest that these abundance patterns 
are not related to the presence of terrestial planets.

\end{abstract} 

  
\section{Introduction\label{secintro}}

The discovery of first exoplanet orbiting a solar-type star by Mayor
\& Queloz (1995) initiated a new and very attractive field in which
the number of studies is continuously increasing. A substantial amount
of spectroscopic data has been collected since then which have allowed 
to perform not only radial velocity planet searches (see e.g Udry \& Santos 
2007; Udry \& Mayor 2008) but also chemical abundance analysis 
(e.g Gonzalez et al. 2001; Sadakane et al. 2002; Ecuvillon et al.
2004; Ecuvillon et al. 2006a; Gilli et al. 2006; Neves et al. 2009) 
and the study of kinematic properties (e.g. Santos et al. 2003; 
Ecuvillon et al. 2007) trying to focus on finding different 
signatures able to distiguish stars with and without known planets.  

The metal-rich nature of star hosting giant planets was firstly 
discussed by Gonzalez (1997, 1998) and later on proved by a uniform 
analysis of large samples of planet-host stars and ``single'' 
(hereafter, ``single'' refers to stars without known planets) stars 
(Santos et al. 2001). 
Further studies has confirmed that the probability of finding a
planet strongly correlates with the metal content of the parent star
(e.g Santos et al. 2004, 2005; Valenti \& Fisher 2005; Sousa et al.
2008). Following these studies and using the same spectroscopic data,
more complete chemical abundance studies have been done and small
differences in some element abundance ratios have been suggested
(Bodaghee et al. 2003; Beir\~ao et al. 2005; Gilli et al. 2006;
Robinson et al. 2006; Neves et al. 2009), although no definite
explanation and/or conclusion has been established yet due to the
contradictory results among these studies. See also the reviews on
chemical abundance trends in planet-host stars by Israelian (2006,
2007, 2008) and Santos (2006, 2008).

Smith et al. (2001) firstly examined the correlations between the
abundance ratio [X/H] as a function of the condensation temperature,
$T_C$, and they found only six stars among a sample of 20 planet-host
stars from Gonzalez et al. (2001) which have positive correlations,
suggesting possible accretion of planetesimals. 
Later on, Ecuvillon et al. (2006b) showed the distributions of
the correlation [X/H] versus $T_C$ in a sample of 88 planet-host stars
and 33 stars without known planets, but they did not find any
significant difference between stars with and without planets.
However, the effective temperature range in this study was probably
too large, $4700 \lesssim T_{\rm eff} {\rm [K]} \lesssim 6400$, 
what may have smoothed out any possible existing signature. Other
studies, but on light elements, have been able to find differences 
between planet hosts and stars without planets (e.g. Israelian et al.
2004; Takeda et al. 2007; Gonzalez et al. 2008). 
For instance, Israelian et al. (2009) found that apparently 
Li is always very depleted in solar twins hosting planets whereas 
this is not the case for similar stars without detected planets. 
This may suggest that planet formation may alter the Li content 
of a star. This result has been confirmed by Sousa et al. 
(2010) and Gonzalez et al. (2010).
 
\begin{deluxetable}{lrrrrrrrr}
\tabletypesize{\scriptsize}
\tablecaption{Spectroscopic observations}
\tablewidth{0pt}
\tablehead{Spectrograph & Telescope & Spectral range &
$\lambda/\delta\lambda$ & Binning & $N_{\rm stars}$\tablenotemark{a} &
$\mid {\rm S/N} \mid$\tablenotemark{b} & 
$\delta \mid {\rm S/N} \mid$\tablenotemark{c} & $\Delta {\rm S/N} $\tablenotemark{d}\\
 & & [{\AA}] & & [{\AA}/pixel] & & & &} 
\startdata
HARPS & 3.6-m & 3800--6900 & 110,000 & 0.010 & 85 & 886 & 450 & 370--2020 \\ 
UVES  & VLT   & 4800--6800 &  85,000 & 0.015 & 8 & 829  & 173 & 550--1100 \\ 
UES   & WHT   & 4150--7950 &  33,000 & 0.040 & 2 & 1400 & --  & -- \\ 
\enddata

\tablecomments{Details of the spectroscopic data used in this work.
The UVES and UES stars are all planet-hosts. 
Two of the UVES stars were observed with a slit of $0.3\arcsec$
providing a resolving power $\lambda/\delta\lambda\sim 110,000$.}

\tablenotetext{a}{Total number of stars including those with and
without known planets.} 

\tablenotetext{b}{Mean signal-to-noise ratio, at $\lambda = 6000$~{\AA}, 
of the spectroscopic data used in this work.} 

\tablenotetext{c}{Standard deviation from the mean signal-to-noise 
ratio.} 

\tablenotetext{d}{Signal-to-noise range of the spectroscopic data.} 

\label{tblobs}      
\end{deluxetable}

In the last decade, many studies have addressed the question whether 
the chemical abundances of the Sun are typical for a solar-type star,
i.e. a star of solar mass and age (e.g. Guftafsson et al. 1998; 
Allende-Prieto et al. 2006). 
The distribution of spectroscopic metallicities of G-type stars 
in the solar neighbourhood has been estimated to be $\sim -0.1$
with a typical dispersion of 0.2~dex (e.g. Edvarsson et al. 1993;
Allende-Prieto et al. 2004). These authors also found small offsets 
in some element abundances with respect to their solar values.
However, these offsets are not always consistent among different
studies which brings some suspicion regarding their existence and 
size. Allende-Prieto et al. (2006) analyzed a sample of solar analogs
and did not find any significant offsets for Si, C, Ca, Ti and Ni, 
suggesting that the offsets reported in previous studies were likely 
the result of systematic errors.  

Recently, Laws \& Gonzalez (2001) performed a differential abundance 
study of the solar twins 16 Cyg A and 16 Gyg B, i.e. two stars with 
stellar parameters and metallicities very close to those of the Sun. 
They used high-quality spectroscopic data of this binary star, 
and found very small differences in their stellar parameters and 
Fe abundances. 
Later on, other authors have proposed and analyzed other solar twins 
using a similar methodology (see Takeda 2005; Mel\'endez et al. 2006,
2007). 
More recently, Mel\'endez et al. (2009) have studied a sample of 11 
solar twins. They also used spectroscopic 
data at high resolving power, $\lambda/\delta\lambda=65,000$, and high 
signal-to-noise ratio (S/N~$\sim 450$ per pixel). They observed
some asteroids to use their spectra as a solar reference. By
performing a fully line-by-line differential analysis, they obtained 
a clear trend, in the mean abundance differences of solar twins with 
respect to the Sun, $\Delta {\rm [X/Fe]_{SUN-TWINS}}$,
as a function of the condensation temperature, $T_C$, of the elements.
The size of this trend is roughly 0.1~dex and goes from carbon at 
$\Delta {\rm [C/Fe]}\sim 0.05$ and $T_C=40$~K to zirconium at $\Delta
{\rm [Zr/Fe]}\sim -0.03$ and $T_C=1736$~K. In spite of this tiny 
signature, they concluded that the most likely 
explanation to this abundance pattern is related to the presence of 
terrestial planets in the solar planetary system.
Ram{\'\i}rez et al. (2009) confirmed this result on a
sample of 64 solar analogs, but with lower signal-to-noise spectra
(S/N~$\sim 200$), although the correlation was not so clear. 


The discovery of more than 400 exoplanets orbiting solar-type stars
detected by the radial velocity technique have provided a substantial
sample of high-quality spectroscopic data, in particular, 
the HARPS GTO planet search program which contains so far about 450 
stars (see e.g. Neves et al. 2009).
The multiplicity of these planetary systems is generating a certain
number of questions about the processes of planet formation and
evolution (see the introduction in Santos et al. 2010).

In the present study, we will use the high-quality   
spectroscopic data to get very accurate chemical abundances in a 
relatively large sample of 95 solar analogs with and without planets, 
and to examine the results reported in Mel\'endez et al. (2009).

\section{Observations\label{secobs}}

In this study we have made use of high-resolution spectroscopic data
obtained with three different telescopes and instruments:
the 3.6-m telescope equipped with HARPS at the {\itshape Observatorio
de La Silla} in Chile, the 8.2-m Kueyen VLT (UT2) telescope equipped 
with UVES at the {\itshape Observatorio Cerro Paranal} in Chile and 
the 4.2-m WHT telescope equipped with UES at the {\itshape 
Observatorio del Roque de los Muchachos} in La Palma, Spain.  
In Table~\ref{tblobs}, we provide the wavelength ranges, resolving
power, S/N and other details of these spectroscopic data. 

The data were reduced in a standard manner, and latter normalized
within IRAF\footnote{IRAF is distributed by 
the National Optical Observatory, which is operated by the Association
of Universities for Research in Astronomy, Inc., under contract with
the National Science Foundation.}, using low-order polynomial fits to
the observed continuum. 

\section{Sample\label{secsample}}

\begin{figure}[!ht]
\centering
\includegraphics[width=5.5cm,angle=90]{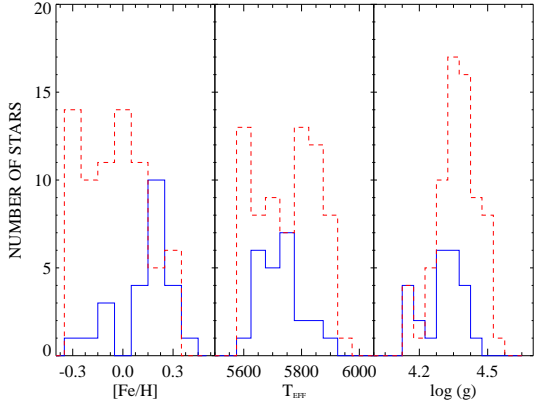}
\caption{Histograms of the stellar parameters and metallicities of the
whole sample of solar analogs hosting planets (solid lines) and
without known planets (dashed lines).}  
\label{fpar}
\end{figure}

To perform a detailed chemical analysis, we selected stars with
S/N~$>350$ from these three spectrographs. We end up with a sample of
24 planet-host stars and 71 ``single'' stars (see Fig.~\ref{fpar}). 
All UVES and UES stars are planet hosts. 
These 95 solar analogs of the sample have stellar
parameters in the ranges $5600 < T_{\rm eff}[{\rm K}] < 5954$ and $
4.0 <\log (g[{\rm cm~s}^{-2}]) < 4.6$, and metallicities in the range
$-0.3 < [{\rm Fe}/{\rm H}] < 0.5$ (see Table~\ref{tblsam}).
In Fig.~\ref{fpar}, we display the histograms of the effective
temperatures, surface gravities and metallicities of the whole sample.
The stars hosting planets are more
metal-rich than the stars without known planets. 
Thus, we defined a new sample of metal-rich 
solar analogs with $0.14 < [{\rm Fe}/{\rm H}] < 0.36$, containing 14
planet hosts and 14 ``single'' stars. We have also considered the case
of solar twins with and without planets. In that case, the number of
stars goes down to only 2 planet-host stars and 5 ``single''
stars. The ranges of the stellar parameters and metallicities are
shown in Table~\ref{tblsam}.

\begin{table}[!ht]
\centering
\scriptsize
\caption[]{Ranges of stellar parameters and metallicities in different
samples\label{tblsam}}
\begin{tabular}{lrrrr}
\noalign{\smallskip}
\noalign{\smallskip}
\noalign{\smallskip}
\hline
\hline
\noalign{\smallskip}
Sample$^{\rm a}$ & $T_{\mathrm{eff}}$ & $\log g$ & $\mathrm{[Fe/H]}$ &
$N_{\rm stars}$$^{\rm b}$ \\
  & [K] & [dex] & [dex] &  \\
\noalign{\smallskip}
\hline
\noalign{\smallskip}
SA   & 5600--5954 & 4.14--4.60 & $-0.30$ -- $+0.50$ & 95 \\
mrSA & 5600--5954 & 4.24--4.60 & $+0.14$ -- $+0.36$ & 28 \\
ST   & 5700--5854 & 4.34--4.54 & $-0.07$ -- $+0.07$ & 7 \\
\noalign{\smallskip}
\hline	     
\hline
\noalign{\smallskip}
\noalign{\smallskip}
\end{tabular}
\begin{minipage}[t]{\columnwidth}

$^{\rm a}\:$ Stellar samples: solar analogs, ``SA'', metal-rich solar
analogs,  ``mrSA'', and solar twins ``ST''.

$^{\rm b}\:$ Total number of stars including those with and
without known planets.

\end{minipage}
\end{table}

\section{Stellar Parameters}

The stellar parameters and metallicities of the whole sample of stars 
were computed using the method described in Sousa et al. (2008), 
based on the equivalent widths (EWs) of 263 \ion{Fe}{1} and 
36~\ion{Fe}{2} lines, measured with the code ARES\footnote{The ARES
code can be downloaded at http://www.astro.up.pt/} (Sousa et al. 2007)
and evaluating the excitation and ionization equilibria. 
The chemical abundance derived for each spectral line was
computed using the 2002 version of the LTE code MOOG (Sneden 1973), 
and a grid of Kurucz ATLAS9 plane-parallel model atmospheres (Kurucz
1993). For the HARPS spectra we just adopted the published results in
Sousa et al. (2008). For the UVES and UES stars, they had
already spectroscopic stellar parameters reported in previous works
(see Santos et al. 2004, 2005) computed from a set of 39 \ion{Fe}{1} 
and 12~\ion{Fe}{2} and using FEROS, SARG and CORALIE spectra. 
Therefore, 
we decided to re-compute the stellar parameters using the higher 
quality data presented in this study and the method described before.
The new stellar parameters are consistent to those in Santos et al. 
(2004, 2005) although the uncertainties are smaller due to the higher
quality of the new data reported in this work.
In Table~\ref{tblpar} we provide the new stellar parameters of the
UVES and UES stars. We note here that the similarities of the
parameters of the stars in the sample with respect to the Sun allow us
to achieve a high accuracy in the chemical analysis since the possible 
uncertainties on the atmospheric model and on the stellar parameters
and abundances are minimized.

\begin{deluxetable}{lrrrrrrrr}
\tabletypesize{\scriptsize}
\tablecaption{Stellar parameters and metallicities from the UVES and UES
spectra}
\tablewidth{0pt}
\tablehead{HD & $T_{\mathrm{eff}}$ & $\delta T_{\mathrm{eff}}$ & $\log
g$ & $\delta\log g$ & $\xi_{\rm t}$ & $\delta\xi_{\rm t}$ & $\mathrm{[Fe/H]}$&
$\delta\mathrm{[Fe/H]}$ \\
 & [K] & [K] & [dex] & [dex] & [\kmso] & [\kmso] & [dex] & [dex]} 
\startdata
106252 & 5880 & 15 & 4.40 & 0.02 & 1.13 & 0.02 & -0.070 & 0.011\\
117207 & 5680 & 29 & 4.34 & 0.04 & 1.06 & 0.04 & 0.250  & 0.022\\
12661\tablenotemark{a}  & 5760 & 37 & 4.33 & 0.07 & 1.09 & 0.04 & 0.385  & 0.028\\
216437 & 5882 & 21 & 4.25 & 0.03 & 1.25 & 0.02 & 0.250  & 0.018\\
217107\tablenotemark{a} & 5679 & 42 & 4.32 & 0.07 & 1.15 & 0.05 & 0.350  & 0.034\\
28185  & 5726 & 40 & 4.45 & 0.05 & 1.08 & 0.05 & 0.230  & 0.031\\
4203   & 5728 & 49 & 4.23 & 0.07 & 1.18 & 0.06 & 0.430  & 0.036\\
70642  & 5732 & 24 & 4.42 & 0.06 & 1.06 & 0.03 & 0.190  & 0.018\\
73526  & 5666 & 25 & 4.17 & 0.04 & 1.12 & 0.03 & 0.260  & 0.020\\
76700  & 5694 & 27 & 4.18 & 0.05 & 1.05 & 0.03 & 0.370  & 0.023\\
\enddata

\tablecomments{Stellar parameters, \teff and \loggl, microturbulent
velocities, $\xi_{\rm t}$, and metallicities , [Fe/H], and their
uncertainties, of the planet-host stars observed with UVES/VLT and 
UES/WHT spectrographs.} 

\tablenotetext{a}{Stars observed with the UES spectrograph at WHT 
telescope.} 

\label{tblpar}      
\end{deluxetable}

\section{Chemical abundances\label{secabun}}

The chemical analysis is done by computing the EWs of spectral
lines using the code ARES (Sousa et al. 2007) for most of the 
elements. We follow the rules given in Sousa et al. (2008) to adjust
the ARES parameter $rejt$ for each spectrum taking into account the S/N
ratio. We fixed the other ARES parameters to: $smoother =4$,
$space=3$, $lineresol= 0.07$, $miniline=2$. However, for O, S and Eu, we 
``manually'' determine the EW by
integrating the line flux; whereas for Zr, 
we ``manually'' performed a gaussian fit taking into account possible 
blends. In both cases, we use the task {\scshape splot} within the 
IRAF package. Finally, the EWs of Sr, Ba and Zn lines were checked 
within IRAF, because, as well as for the elements O, S, Eu and Zr, 
for some stars, ARES did not find a good fit due to numerical 
problems and/or bad continuum location. 
For further details see Sect.~\ref{secsun}.

Once the EWs are measured, we use the LTE code MOOG (Sneden 1973) to
compute the chemical abundance provided by each spectral line, using
the appropriate ATLAS model atmosphere of each star. We determine the 
mean abundance of each element relative to its solar abundance (see
Sect.~\ref{secsun}) by computing the line-by-line mean difference. 
However, to avoid problems with wrong EW measurements of some 
spectral lines, we rejected all the lines with an abundance 
different from the mean abundance by more than a factor of 1.5 
times the rms. We also checked that we get the same results when 
using a factor of 2 times the rms, but we decided to stay in a
restrictive position. 
The line-by-line scatter in the differential abundances goes on 
average, for the 5 ``single'' solar twins (see
Sect.~\ref{sectwins}), from $\sigma \sim 0.012-0.014$ for elements 
like Ni, Si and Cr, to $\sigma \sim 0.048-0.056$ for Mg, Zn and Na, 
whereas for the whole sample 
of 71 solar analogs without planets, the line-by-line scatter goes 
on average from $\sigma \sim 0.016$ for elements like Ni, Si and Cr, 
to $\sigma \sim 0.050$ for Mg and Na and to $\sigma \sim 0.074$ for Zn.

\subsection{Atomic data\label{secatom}}

The oscilator strengths of spectral lines used in this study were
extracted from three different works. The line data from Na ($Z=11$) 
to Ni ($Z=28$) were compiled from Neves et al. (2009). Among
these elements, the only element with abundance derived from
an ionized state is Sc. \ion{C}{1} and \ion{S}{1} data is based on
Ecuvillon et al. (2004), and [\ion{O}{1}] from Ecuvillon et al.
(2006a). We note that \ion{S}{1} multiplet lines were combined into a
single line by adding the $gf$ values. The three \ion{Zn}{1} were
extracted from Ecuvillon et al. (2004) and Reddy et al. (2003), as
well as the four \ion{Cu}{1} lines. However, three \ion{Cu}{1} lines were
rejected high dispersion in the resulting Cu abundances probably due
to blending effects with other element lines. 
The atomic data of the s-process elements Sr, Ba,
Y, Zr, Ce, Nd and the r-process element Eu were also extracted from 
Reddy et al. (2003). 
Similarly, one \ion{Ba}{1} line was discarded since its too 
high strength provided very different abundance from the other two
lines (see Sect.~\ref{secsun}).

We perform a completely differential analysis to the
solar abundances on a line-by-line basis, and therefore, the
uncertainties on the oscilator strengths are nearly irrelevant, since
most of the lines are in the linear part of the curve of growth, with
the exception of some Ba, Fe, and probably also some Zn, Mn, and Ca 
lines.  

\subsection{The solar reference\label{secsun}}

This fully differential analysis is, at least, internally consistent. 
For this reason, we have used two 
HARPS solar spectra\footnote{The HARPS solar spectra can be downloaded at 
http://www.eso.org/sci/facilities/lasilla/instruments
/harps/inst/monitoring/sun.html} as solar reference: a daytime sky
spectrum (Dall et al. 2006) with a S/N~$\sim 1000-1200$ and the 
spectrum of the \emph{Ganymede}, a \emph{Jupiter}'s satellite, 
with a S/N~$\sim 350-400$. However, it should be mentioned that the
stars 
We note here that the spectral lines in solar spectra obtained on the
daytime sky are known to exhibit EW and line depth changes (e.g. Gray
et al. 2000). Therefore, although the asteroid spectrum has a lower
S/N ratio than the sky spectrum, we think it is more likely a better
reference solar spectrum and we have adopted the asteroid element
abundances as our solar reference in this work. However, we will
compare the results obtained with the sky spectrum when analysing the
solar twins (see Sect.~\ref{sectwins}), just to show how important is
to have a good solar reference spectrum.
It should be mentioned, however, that the UVES and UES stars
were also analysed using the HARPS \emph{Ganymede} spectrum as solar
reference due to the lack of an appropriate solar spectrum observed
with these two instruments.

\begin{figure}[!ht]
\centering
\includegraphics[width=5.5cm,angle=90]{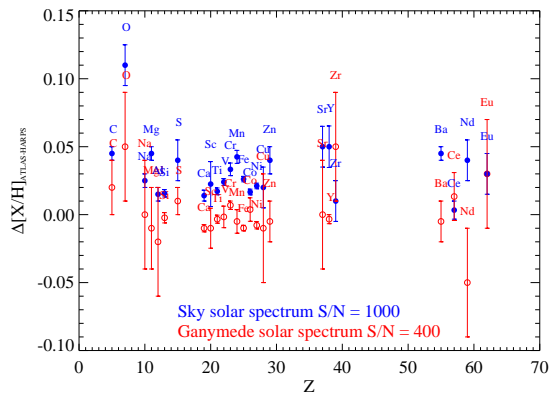}
\caption{Abundance difference on a line-by-line basis between the
Kurucz atlas solar spectrum and, a daytime sky spectrum (filled circles)
and a \emph{Ganymede} spectrum (open circles). The
error bars show the standard deviation of the line abundance
differences divided by the square root of the number of lines of each
element.}  
\label{fsun}
\end{figure}

In Fig.~\ref{fsun} we display the difference between the
solar chemical abundances measured in the Kurucz atlas solar spectrum
(Kurucz et al. 1984) and the sky and Ganymede solar HARPS spectra, 
computed in the same way as the chemical abundances of the whole 
sample of solar analogs. 
The abundances of the elements O, Sr, Cu, Eu, Nd, 
and Zr were determined from only one spectral line; those of the
elements C, S, Na, Mg, Al and Ba from 2 lines; those of Zn, Y, Ce 
from three lines; and the abundances of the rest of elements in 
Fig.~\ref{fsun} from more than 5 lines.
In Fig.~\ref{fsun} we adopt an uncertainty of 0.04~dex and
0.015~dex for the asteroid and sky spectrum, respectively, for the
elements with only one line. 
For the other elements, the adopted uncertainty is computed
from the standard deviations of the individual line measurements
divided by the square root of the number of lines. 
Most of the element abundance differences with respect to the Sky 
HARPS spectrum are slightly higher than zero, whereas these 
abundance differences are roughly zero in the case of the Ganymede
HARPS spectrum. This is probably related to the EW variations
in the sky daytime solar HARPS spectrum. In general, elements with
relatively small line equivalent widths ($EW\lesssim 10$~m{\AA}) show the 
larger discrepancies with the solar atlas abundances. 
We list below some detailed information on the most {\it problematic}
elements: 

\begin{itemize}

\item[a)] Oxygen abundance is derived from the forbidden [\ion{O}{1}] line at
6300.304~{\AA} (with a expected EW of $\sim 4.1$~m{\AA}) and is 
severely blended with the [\ion{Ni}{1}]~$\lambda$~6300.34~{\AA} (with a 
expected EW of $\sim 1.3$~m{\AA}, e.g. Nissen et al. 2002). In the
atlas solar spectrum the whole feature has an EW of $\sim 5.4$~m{\AA}
which provides an abundance, A(O)\footnote{A(X)$=\log [N({\rm
X})/N({\rm H})]+12$}, of 8.74~dex. It is the element with the smaller
EW measurement in this study. 

\item[b)] Zirconium abundances come from the line
\ion{Zr}{2}~$\lambda$~5112.28~{\AA} which slightly blended with 
two weaker lines. We estimate an EW of $\sim 9.5$~m{\AA} in the solar
atlas spectrum. In the HARPS spectrum of the asteroid, the \ion{Zr}{2}
abundance is substantially different from that of the atlas
spectrum. In addition, the sky and the asteroid spectra  
show different EW, $\sim$~9.2~m{\AA} and 8.6~m{\AA}, respectively, 
which translates into an abundance difference of $\sim 0.05$~dex.
This does not seems to be related with the S/N ratio, which provides 
an uncertainty on the EW measurement of $\sim$~0.05~m{\AA} and 
0.15~m{\AA}, respectively, from the Cayrel's formula (Cayrel 1988).

\item[c)] Neodymium is estimated from
the line \ion{Nd}{2}~$\lambda$~5092.80~{\AA} which is relatively
isolated or weakly blended. However, in this case the sky spectrum 
clearly show smaller EW than the asteroid spectra, 
$\sim$~7.6~m{\AA} and 9.2~m{\AA}, respectively, resulting  
into an abundance difference of $\sim 0.1$~dex. 

\item[d)] Europium is determined from the line
\ion{Eu}{2}~$\lambda$~6645.13~{\AA}, and is also blended with two
lines of smaller EWs. The solar atlas spectrum shows EW of $\sim
6.4$~m{\AA}.

\begin{figure*}[!ht]
\centering
\includegraphics[width=5.5cm,angle=90]{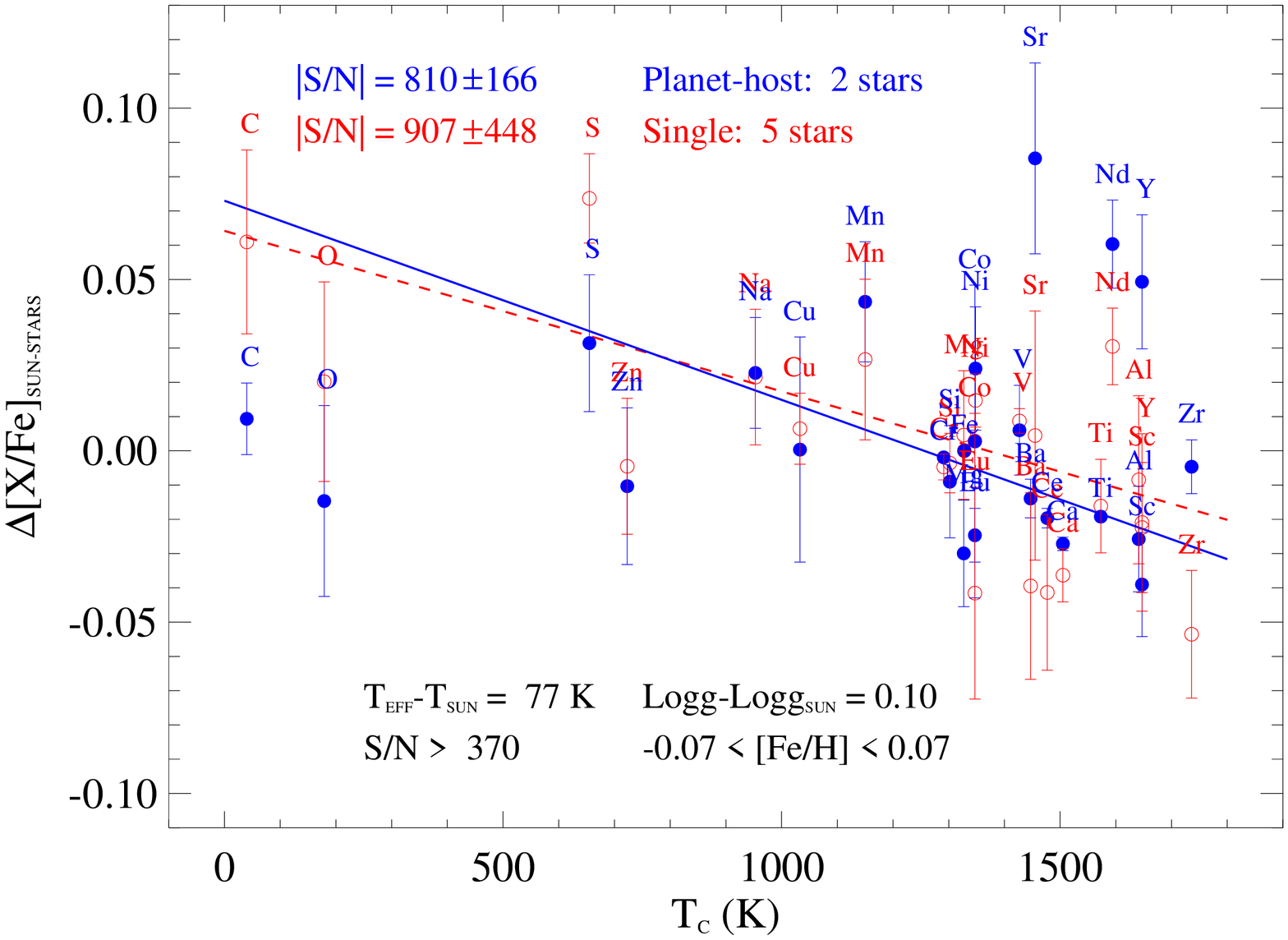}
\includegraphics[width=5.5cm,angle=90]{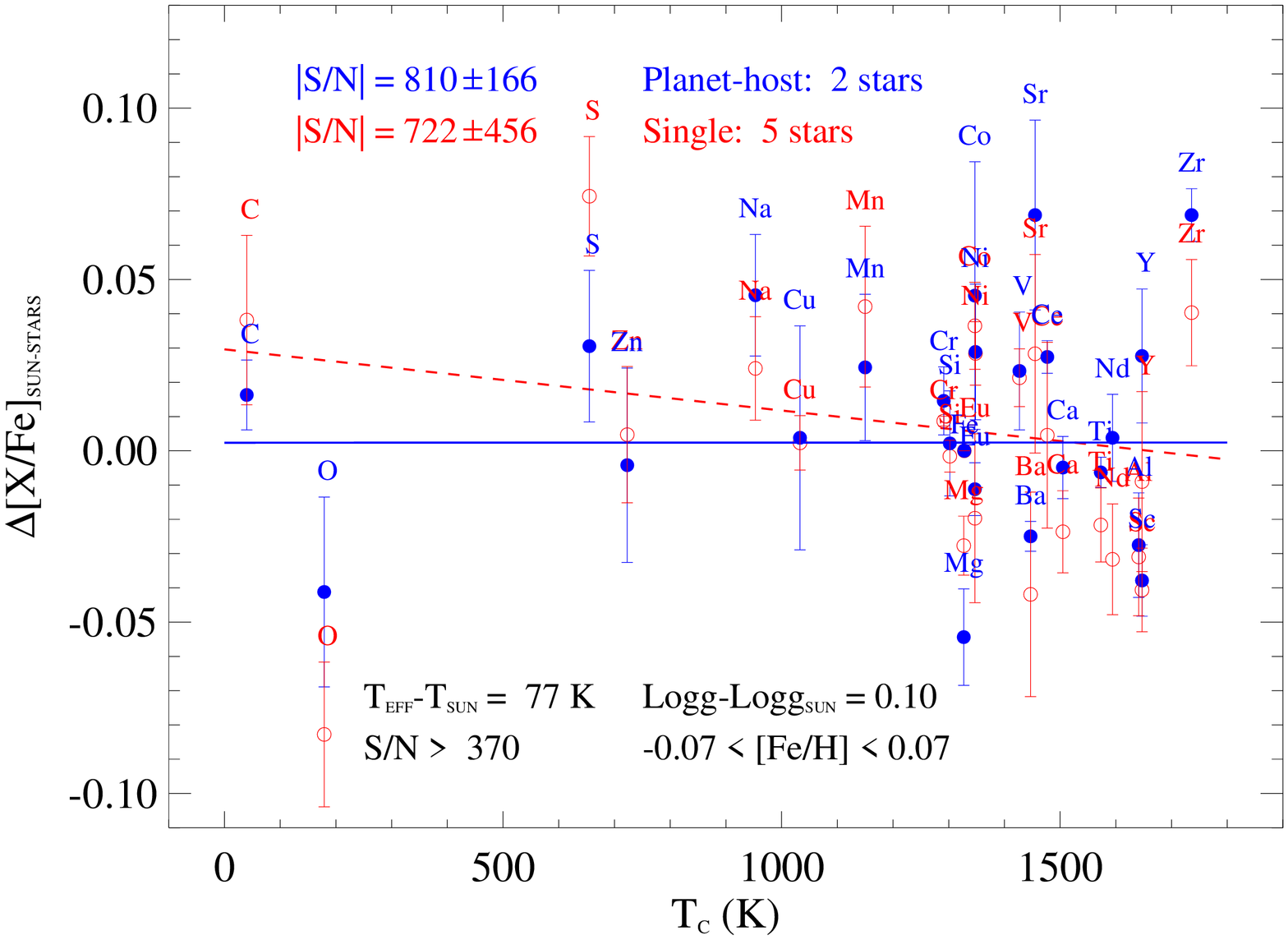}
\caption{Mean abundance differences, $\Delta {\rm [X/Fe]_{SUN-STARS}}$, 
between the Sun, and 2 planet hosts (filled circles) and 5 ``single''
stars (open circles). Error bars are the standard deviation from the
mean divided by the square root of the number of stars. Linear fits to
the data points weighted with the error bars are also displayed for
planet hosts (solid line) and ``single'' stars (dashed line). The left
panel shows the results using the solar spectrum of the
\emph{Ganymede} and the right panel, using the sky solar 
spectrum taken in daytime. Note that the average S/N ratios are
different in both panel for ``single'' stars. This is due to the fact
that different stars fulfill the [Fe/H] condition of solar twins (see
Sect.~\ref{sectwins}).}
\label{ftwin}
\end{figure*}

\item[e)] Barium, which comes from the two strong lines of
\ion{Ba}{2}~$\lambda$~5853.69, and~$\lambda$~6496.91~{\AA}, can be
easily measured, in principle, although the
[\ion{Ba}{2}/Fe] versus [Fe/H] trend displays the largest dispersion 
among the elements analyzed in this work (see Sect.~\ref{secgal}). 
This may be related to the dependence of the strong \ion{Ba}{2} 
lines on the microturbulence and other factors.
 
\item[f)] Cerium is derived from three lines of
\ion{Ce}{2} at~4523.08, 4628.16, and 4773.96~{\AA}. The error bars
are relatively small but it is also off from the zero level in
Fig.~\ref{fsun}. 

\item[g)] Strontium and Yttrium abundances are estimated 
from the relatively strong lines \ion{Sr}{1}~$\lambda$~4607.34~{\AA}
and the three lines \ion{Y}{2} at~5087.43, 5200.42, and
5402.78~{\AA}. As Zr, Ba and Ce, these two elements also show a
large dispersion in their Galactic chemical evolution trends 
(see Sect.~\ref{secgal}).
 
\end{itemize}

\section{Discussion\label{secdisc}}

In this section we will discuss the abundance ratios of different
elements as a function of the metallicity, [Fe/H], as well as 
the relation, for different samples of solar analogs, between 
the abundance difference, $\Delta {\rm [X/Fe]_{SUN-STARS}}$ and 
the 50\% equilibrium condensation temperature, $T_C$, for
a solar-photosphere composition gas with 50\% element condensation, 
which is the temperature at which half of an element
is kept in the gas and the other half is kept into condensates
(Lodders 2003).    

\subsection{Galactic abundance trends\label{secgal}}

\begin{figure*}[ht]
\centering
\includegraphics[width=12.5cm,angle=90]{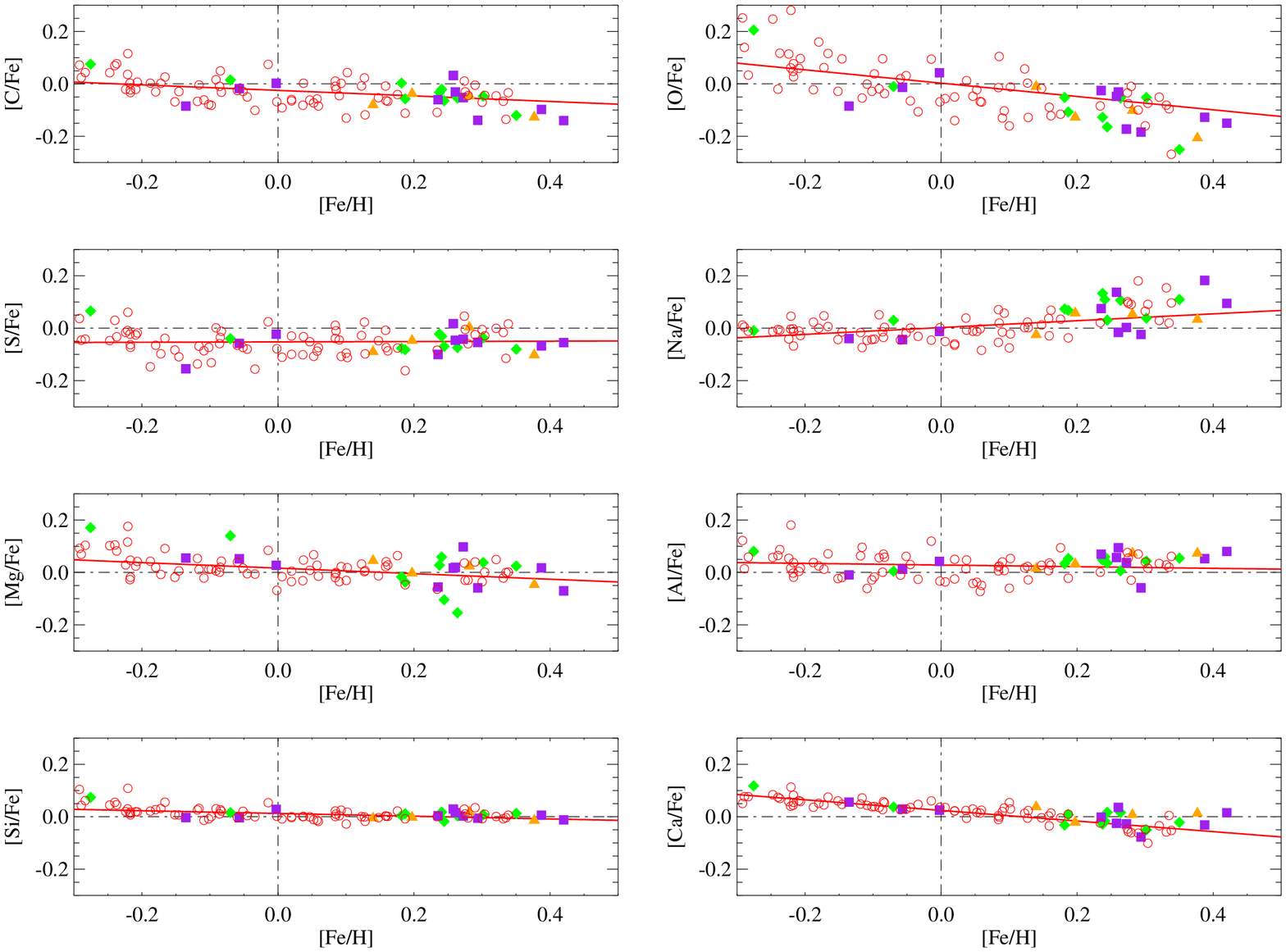}
\caption{Chemical abundance ratios [X/Fe] versus [Fe/H] for the whole
sample of solar analogs, containing 71 ``single'' stars (open
circles) and 24 stars hosting planets, with the most massive planet 
in an orbital period $P_{\rm orb} < 25$ 
days (filled diamonds), $150 < P_{\rm orb} < 650$ days (filled 
squares), $1000 < P_{\rm orb} < 4300$ days (filled triangles).  
Solid lines shows linear fits to the data points of the 
``single'' stars.}  
\label{fgala}
\end{figure*}

\begin{figure*}[ht]
\centering
\includegraphics[width=12.5cm,angle=90]{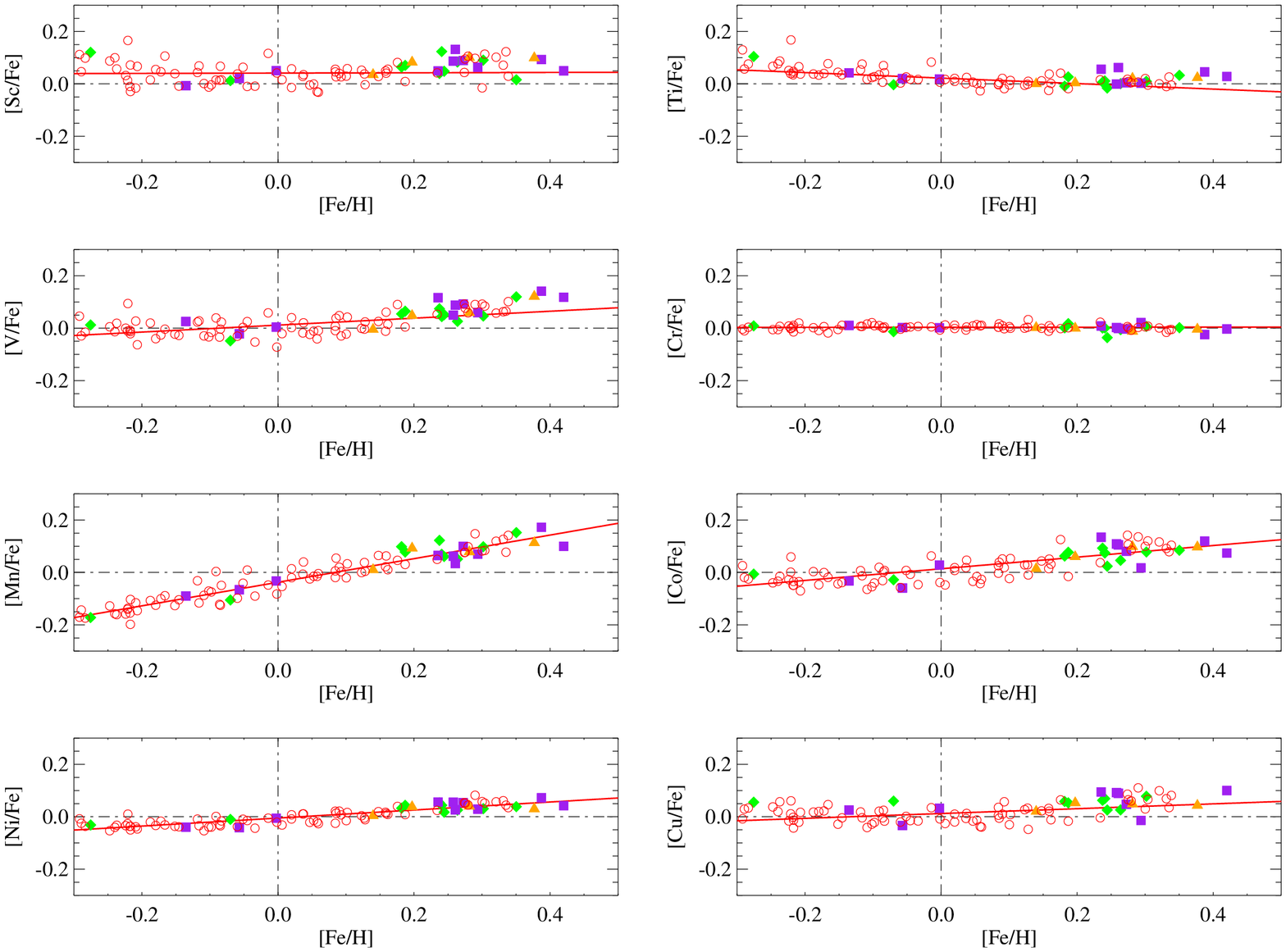}
\caption{Same as Fig.~\ref{fgala} but for other elements.}  
\label{fgalb}
\end{figure*}

\begin{figure*}[ht]
\centering
\includegraphics[width=12.5cm,angle=90]{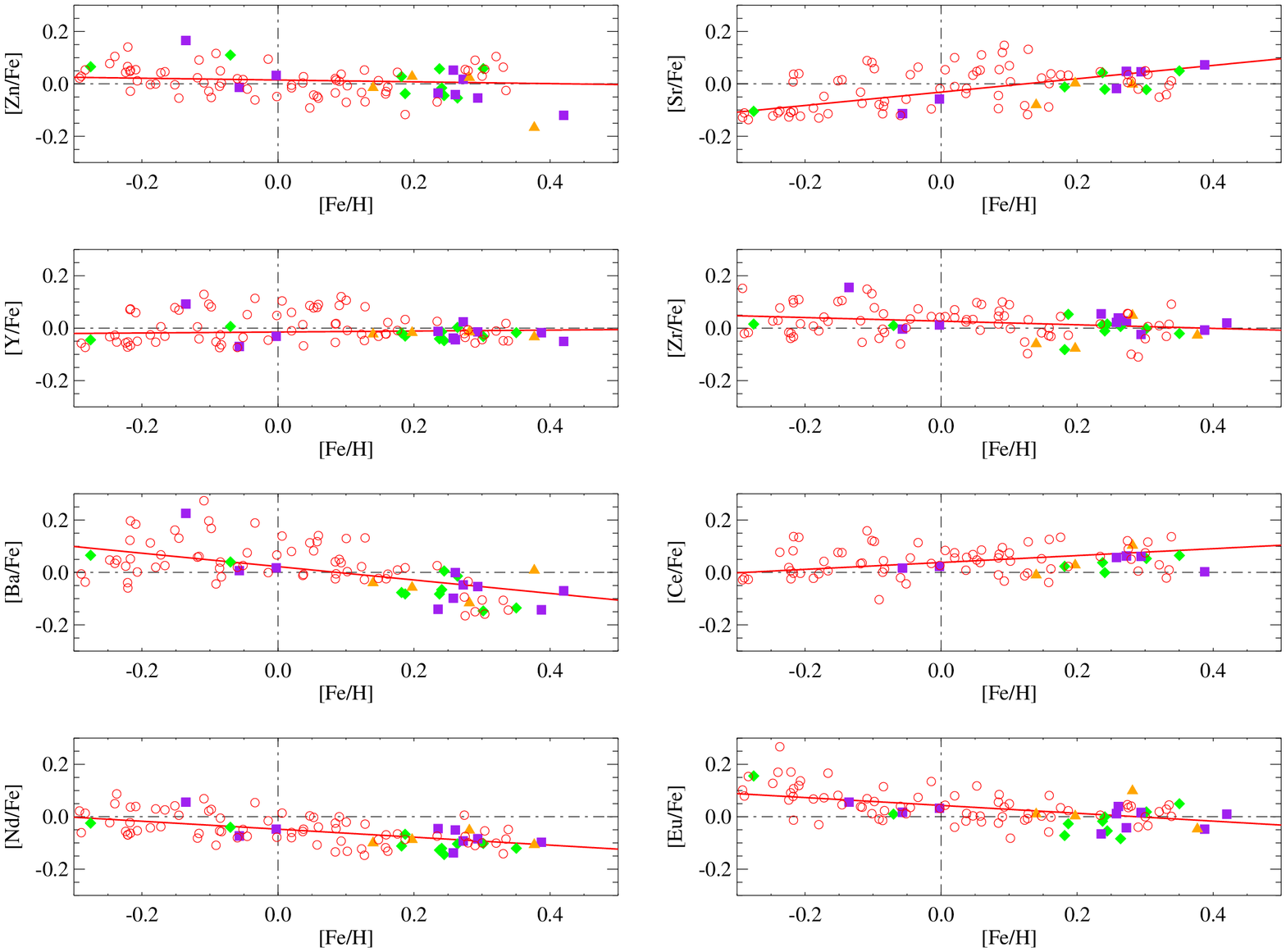}
\caption{Same as Fig.~\ref{fgala} but for other elements.}  
\label{fgalc}
\end{figure*}

The high-quality of HARPS and UVES data presented in this work allows
us to derive very accurate abundance ratios of many elements. 
In Figs.~\ref{fgala},~\ref{fgalb} and~\ref{fgalc} (available online),
we show the abundance ratios [X/Fe] versus [Fe/H] for the 
whole sample of solar analogs analyzed in this work. These trends are
compatible with those in previous works (e.g. Bodaghee et al. 2003; 
Beir\~ao et al. 2005; Bensby et al. 2005; Gilli et al. 2006;
Reddy et al. 2006; Takeda 2007; Neves et al. 2009; Ram{\'\i}rez
et al. 2009). We note here the low dispersion of most of the elements
analyzed in this work, with the exception of some elements discussed 
in Sect.~\ref{secsun}. In Table~\ref{tabplaex} we provide an
example of the tables containing all the element abundance ratios
[X/Fe] which are all available online.

\begin{table*}
\centering
\scriptsize
\caption{Abundance ratios [X/Fe] of solar analogs without known planets$^{\rm a}$\label{tabplaex}}
\begin{tabular}{lrrrrrr}
\noalign{\smallskip}
\noalign{\smallskip}
\noalign{\smallskip}
\hline
\hline
\noalign{\medskip} 
HD       &     [C/Fe]      &     [O/Fe]      &     [S/Fe]      &     [Na/Fe]      &     [Mg/Fe]      &     [Al/Fe]      \\
\noalign{\medskip} 
\hline
\hline
\noalign{\medskip} 
   10180 & $ 0.011\pm0.035$ & $-0.014\pm0.070$ & $ 0.011\pm0.021$ & $ 0.071\pm0.049$ & $-0.019\pm0.078$ & $ 0.006\pm0.070$ \\
  102365 & $ 0.072\pm0.113$ & $ 0.252\pm0.070$ & $ 0.037\pm0.134$ & $ 0.012\pm0.057$ & $ 0.092\pm0.014$ & $ 0.122\pm0.014$ \\
  104982 & $ 0.003\pm0.021$ & $-0.022\pm0.070$ & $-0.147\pm0.134$ & $ 0.023\pm0.049$ & $ 0.043\pm0.021$ & $ 0.038\pm0.042$ \\
  106116 & $-0.053\pm0.021$ & $-0.018\pm0.070$ & $-0.048\pm0.014$ & $ 0.012\pm0.071$ & $-0.013\pm0.064$ & $ 0.027\pm0.035$ \\
  108309 & $ 0.023\pm0.042$ & $ 0.003\pm0.070$ & $ 0.023\pm0.071$ & $ 0.023\pm0.085$ & $ 0.003\pm0.071$ & $ 0.023\pm0.028$ \\
  109409 & $-0.066\pm0.021$ & $-0.081\pm0.070$ & $-0.016\pm0.021$ & $ 0.154\pm0.049$ & $-0.021\pm0.127$ & $ 0.029\pm0.028$ \\
  111031 & $-0.045\pm0.021$ & $-0.009\pm0.070$ & $-0.060\pm0.014$ & $ 0.090\pm0.014$ & $-0.030\pm0.099$ & $-0.030\pm0.071$ \\
...& & & & & & \\
\noalign{\medskip} 
\hline
\noalign{\smallskip}
\noalign{\smallskip}
\end{tabular}
\begin{minipage}[t]{\textwidth}

$^{\rm a}\:$ Tables~\ref{tabpla1},~\ref{tabpla2},~\ref{tabpla3},
\ref{tabpla4},~\ref{tabsin1},~\ref{tabsin2},~\ref{tabsin3},
\ref{tabsin4} are published in its entirety in the electronic 
edition of the Astrophysical Journal. A portion is shown here for 
guidance regarding its form and content.

\end{minipage}
\end{table*}

\subsection{HARPS solar twins\label{sectwins}}

We examined the whole sample of HARPS targets trying to search for 
solar twins defined in the same way as in Mel\'endez et al. (2009)
and we found 2 planet host and 5 ``single'' stars 
(see Table~\ref{tblsam}). In Fig.~\ref{ftwin} we display the mean
abundance difference, $\Delta {\rm [X/Fe]_{SUN-STARS}}$, versus the
condensation temperature, $T_C$, for both reference solar spectra,
the \emph{Ganymede} spectrum (left panel) and the 
daytime sky spectrum (right panel). The main differences between both
panels are for the elements O, Nd, Zr, and Y which were discussed in
Sect.~\ref{secsun}. Although the error bars are relatively large to
search for any clear trend, it seems that the mean abundance ratios of 
refratories are on average smaller than those of volatiles. In this
plot we also show linear fits to the data points weighted by their error
bars. These fits show the decreasing trend of the mean abundance 
ratios with condensation temperature already reported in 
Mel\'endez et al. (2009) but the trend is not so clear. We note here
that the S/N is greater than 500 for all stars in this figure, 
except one ``single'' star with a S/N~$\sim 370$. 
We display on the upper-right corner 
of both panels of Fig.~\ref{ftwin} the mean S/N and its standard 
deviation. Besides, the mean S/N ratios are different in both panel 
for ``single'' stars. This is due to the fact
that just one star in each panel did not fulfill the [Fe/H] condition 
of solar twins. 
This is because small offsets in the range 0.01-0.05 dex are found in
the differential Fe abundances and probably different Fe lines were
used for each star in each case (see Sect.~\ref{secabun}). 
In the right panel of Fig.~\ref{ftwin}
the fits change completely mainly due to the position of O and Zr 
abundance ratios. 
We have checked that expanding the metallicity range up to
$\pm 0.15$~dex only increases the number ``single'' stars, up to 11,
which is the same number of stars as in Mel\'endez et al. (2009).
However, this does not change significantly either the position of the
data points or the slope of the linear fit.

We find some slight differences between the abundance
ratios of stars hosting planets and ``single'' stars. However, the
fits have similar slopes, but only in the left panel of
Fig.~\ref{ftwin}. In addition, the number of planet hosts is too small
to allow us to make a strong statement on implications of these
differences.

\begin{figure}[!ht]
\centering
\includegraphics[width=5.5cm,angle=90]{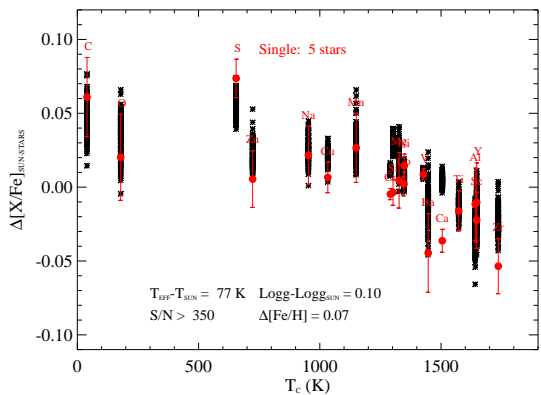}
\caption{100 Monte Carlo simulations of the mean element abundance ratios 
(asterisks) reported in Mel\'endez et al.(2009) taking into account 
the error bars shown in Fig.~\ref{ftwin}, and in comparison with 
the results (filled circles) shown in Fig.~\ref{ftwin}.}  
\label{fsim1}
\end{figure}

We have tried to understand why our results do not look like those
of Mel\'endez et al. (2009). Our data has better S/N and resolving
power than those of Mel\'endez et al. but it seems that our dispersion
in the mean abundance ratios versus $T_C$ and also around the linear 
fits is larger than in Mel\'endez et al. (2009). The
number of solar twins in our study is smaller, although we think 
this may not explain the differences between both studies. 
We have made two
Monte Carlo simulations to check if our results are consistent with 
those of Mel\'endez et al. (2009), according to our error bars. 

\begin{figure}[!ht]
\centering
\includegraphics[width=5.5cm,angle=90]{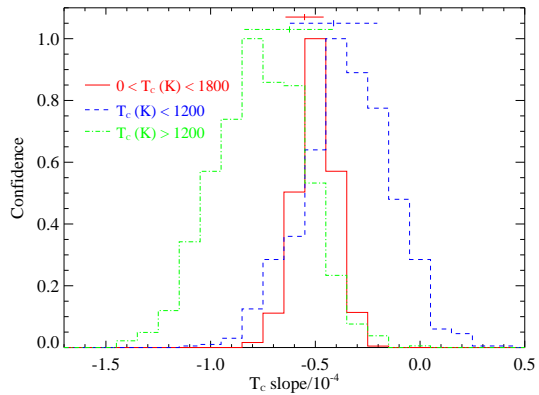}
\caption{Histograms of slopes of linear fits to 1000 simulations 
of the points on the fits to mean element abundance ratios reported 
in Mel\'endez et al. (2009). The histograms show the distributions for
$T_C {\rm [K]}< 1200$ (dashed line), $T_C{\rm [K]} > 1200$
(dashed-dotted line), and $0 < T_C {\rm [K]}< 1800$ (solid line). On
the top of this figure, we display the slopes and error bars of our
results for the 5 ``single'' twins stars shown in Fig.~\ref{ftwin}.
See also Table~\ref{tblslope}.} 
\label{fsim2}
\end{figure}

In the first simulation (see Fig.~\ref{fsim1}), we randomly generate 
5 abundance ratios (corresponding to the 5 solar twins) using 
gaussian distributions around the mean abundance ratios reported 
in Mel\'endez et al. (2009) taking into account the error bars 
displayed in the left panel of Fig.~\ref{ftwin}. 
We then compute the mean abundance ratios from these 5 points and
finally displayed 100 simulations for each element in Fig.~\ref{fsim1}
in comparison with our results of the 5 ``single'' stars 
(see left panel of Fig.~\ref{ftwin}). 
Note that Mel\'endez et al. (2009) do not include in their analysis
the elements Ce, Nd, Sr and Eu, which are in fact some of the outliers
in Fig.~\ref{fsun} (see also Sect.~\ref{secsun}), although they 
include N, P, and K. 
This simulation shows that our results and those in
Mel\'endez et al. are indeed consistent, although in some cases only
marginally, for most of the elements, according to the error bars 
shown in Fig.~\ref{ftwin}, with the
exception of Si, Ca and Cu. The line-by line scatter of Ca and Si 
differential abundances for the 5 ``single'' stars are on average 
0.013 and 0.018~dex whereas the star-to-star scatter, 0.022 and
0.019~dex, respectively. The star-to-star scatter for Cu is 0.027~dex.
Note that the error bars in Fig.~\ref{ftwin}
contain the star-to-star scatter divided by the square root of 
the number of stars, but only one line of Cu was used in the abundance
computation (see Sect.~\ref{secatom})
The fact that the star-to-star and line-by-line scatters are roughly 
equal and so small may exacerbate the disagreement between our Ca and
Si mean abundances and those in Mel\'endez et al. (2009). 
In addition, the scatter of our mean element abundance ratios around 
the linear fits is
higher than in Mel\'endez et al., which may be related to the solar
reference HARPS spectrum used in this analysis.

\begin{figure}[!ht]
\centering
\includegraphics[width=5.5cm,angle=90]{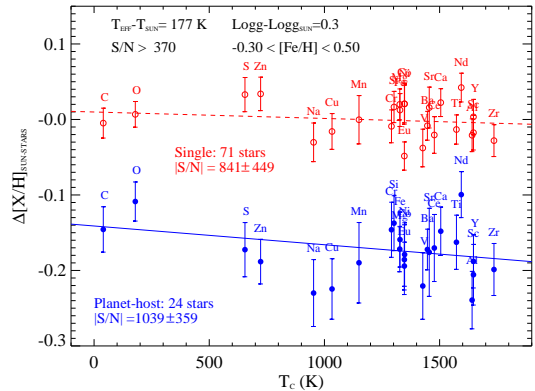}
\caption{Same as in Fig.~\ref{ftwin} but for the mean ratio 
$\Delta {\rm [X/H]_{SUN-STARS}}$ in the 24 planet host and 
71 ``single'' stars of the whole sample of solar analogs, 
using the \emph{Ganymede} spectrum as solar reference.}  
\label{fall}
\end{figure}

\begin{figure*}[!ht]
\centering
\includegraphics[width=10cm,angle=90]{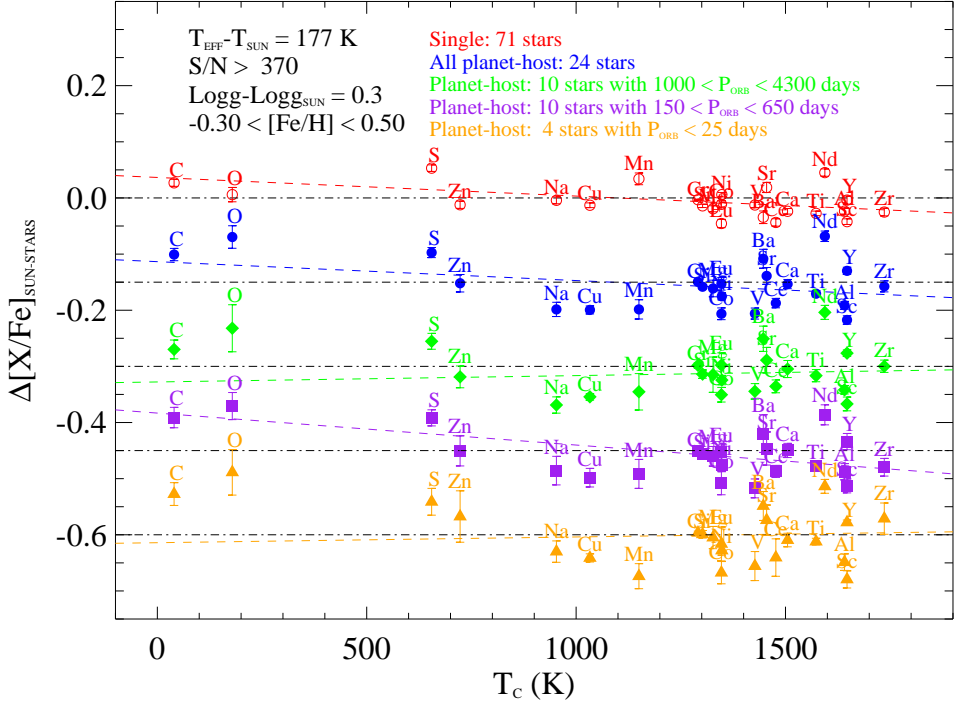}
\caption{Same as in Fig.~\ref{ftwin} containing 71 ``single'' stars 
(open circles) and 24 stars hosting planets (filled circles),
with the most massive planet in an orbital period $P_{\rm orb} < 25$ 
days (4 stars, filled diamonds), $150 < P_{\rm orb} < 650$ days 
(10 stars, filled squares), $1000 < P_{\rm orb} < 4300$ days 
(10 stars, filled triangles). The $\Delta {\rm [X/H]_{SUN-STARS}}$ 
and linear fits has been artificially shifted by $-0.15$~dex, for the 
sake of clarity. Horizontal dashed-dot lines show the zero point 
levels for each set of points.}
\label{fallpar}
\end{figure*}

In the second simulation, we derive a linear fit to the mean abundance
ratios presented in Mel\'endez et al. (2009), and define, as new
reference abundance ratios, the points on this fit at the condensation
temperature of each element. Then we generate, around these new
points, abundance ratios using again gaussian distributions which take
into account the error bars of the results of ``single'' stars 
(see Fig.~\ref{ftwin}).
Finally, for each simulation we perform a linear fit 
to these abundance ratios.
We repeat the same procedure only for elements with $T_C$ above or
below 1200~K. Three histograms of 1000 simulations are displayed in 
Fig.~\ref{fsim2}. The slopes of the linear fits coming from our 
data are shown on the top of this figure (see also
Table~\ref{tblslope}) and seems to be clearly consistent with 
the peak of these histograms, although for $T_C {\rm [K]}> 1200$,
our point slightly moves to higher values of the slopes. 
In Table~\ref{tblslope} we provide the values of the slopes in
the linear fits to the mean abundance ratios $\Delta {\rm
[X/Fe]_{SUN-STARS}}$ as a function of the condensation temperatures
$T_C$, at different ranges of $T_C$.

\subsection{All solar analogs\label{secall}}

\begin{deluxetable}{lrrrr}
\tabletypesize{\scriptsize}
\tablecaption{Slopes of the linear fit to the mean values $\Delta {\rm
[X/Fe]_{SUN-STARS}}$ versus $T_C$} 
\tablewidth{0pt}
\tablehead{Sample\tablenotemark{a} & $0 <T_{C}[{\rm K}] < 1800$ & 
$T_{C}[{\rm K}] < 1200$ & $T_{C}[{\rm K}] > 1200$ & 
$N_{\rm stars}$\tablenotemark{b}}  
\startdata
sSA   & $-0.356\pm0.032$ & $-0.450\pm0.067$ & $-0.468\pm0.070$ & 71 \\
pSA   & $-0.327\pm0.051$ & $-1.189\pm0.116$ & $-0.281\pm0.104$ & 24 \\
smrSA & $-0.235\pm0.058$ & $-1.285\pm0.119$ & $ 0.179\pm0.114$ & 14 \\
pmrSA & $-0.158\pm0.051$ & $-1.596\pm0.107$ & $ 0.327\pm0.104$ & 14 \\
sST   & $-0.468\pm0.090$ & $-0.513\pm0.214$ & $-0.306\pm0.202$ & 5 \\
pST   & $-0.581\pm0.025$ & $ 0.211\pm0.147$ & $-0.642\pm0.027$ & 2 \\
PHs   & $ 0.113\pm0.069$ & $-1.112\pm0.160$ & $ 0.154\pm0.144$ & 4 \\
PHm   & $-0.569\pm0.082$ & $-1.112\pm0.197$ & $-0.619\pm0.167$ & 10 \\
PHl   & $ 0.102\pm0.077$ & $-1.137\pm0.192$ & $ 0.299\pm0.123$ & 10 \\
\enddata

\tablecomments{Slopes of the linear fit of the mean abundance ratios,
$\Delta {\rm [X/Fe]_{SUN-STARS}}$, as a function of the condensation
temperature, $T_C$, using the elements with $T_C$ within the 
specified interval, for different stellar samples.} 

\tablenotetext{a}{Stellar samples: ``single'' solar analogs, ``sSA'',
planet-host solar analogs, ``pSA'', ``single'' metal-rich solar
analogs,  ``smrSA'', planet-host metal-rich solar analogs, ``pmrSA'',
 ``single'' solar twins ``sST'', planet-host solar twins ``pST'', 
planet-host solar analogs with the most massive planet in an 
orbital period $P_{\rm orb} < 25$ days, ``PHs'', 
$150 < P_{\rm orb} < 650$ days , ``PHm'', 
$1000 < P_{\rm orb} < 4300$ days, ``PHl''.} 

\tablenotetext{b}{Total number of stars including those with and
without known planets.}

\label{tblslope}      
\end{deluxetable}

The whole sample of solar analogs presented in this paper contains 24
planet hosts and 71 stars without known planets. In Fig.~\ref{fall} we
display $\Delta {\rm [X/H]_{SUN-STARS}}$ of the sample versus 
$T_C$, only using the \emph{Ganymede} spectrum as 
solar reference. The abundance pattern in this figure is very similar 
to that of $\Delta {\rm [X/Fe]_{SUN-STARS}}$ (see Fig.~\ref{fallpar}), 
although due to 
higher mean metallicity of the stars hosting planets, their mean 
abundance ratios appear shifted. In addition, the error bars in 
this plot are larger than in a plot $\Delta {\rm [X/Fe]_{SUN-STARS}}$,
since the dispersion in [X/H] is typically larger than in [X/Fe] due
to chemical evolution effects. Those elements, like Mn and
O, with steeper trends [X/Fe] versus [Fe/H], will show larger 
abundance differences in the plot $\Delta {\rm [X/Fe]_{SUN-STARS}}$ 
(see Fig.~\ref{fallpar}). 
However, in the plot $\Delta {\rm [X/H]_{SUN-STARS}}$ these differences
are relatively smaller and most points seems to agree well with the 
linear fits. Both stars with and without planets show a similar 
decreasing trends towards increasing values of $T_C$. 
Nearly equal results are found for the plot $\Delta {\rm
[X/Fe]_{SUN-STARS}}$ versus $T_C$. 

\begin{figure}[!ht]
\centering
\includegraphics[width=5.5cm,angle=90]{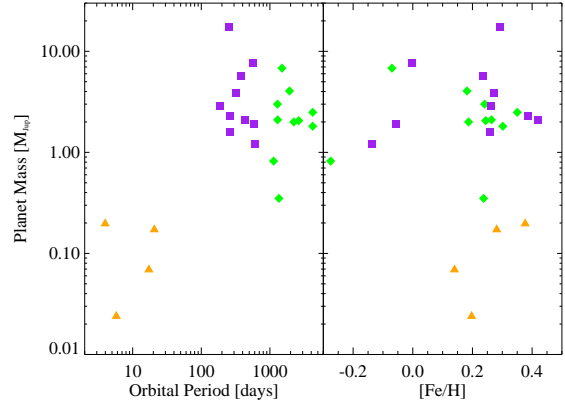}
\caption{Mass of the most massive planet of each star versus the
orbital period of the planet (left panel) and metallicity [Fe/H]
(right panel). The symbols as in Fig.~\ref{fallpar}.}  
\label{fallmo}
\end{figure}

The stars hosting planets studied in this work shows a variety 
of planetary systems with
different orbital periods, $P_{\rm orb}$, from 3 to 4200~days, and
minimum masses, i.e. $M_p \sin i$, from 0.02 to 17.4~$M_{\rm Jup}$, 
where $M_p$ is the mass of the primary planet, i.e. the most massive 
planet in the planetary system, $M_{\rm Jup}$, the Jupiter mass, and 
$i$, the orbital inclination. Among the 24 planet-host stars in the
sample, six are planetary systems containing two planets, 
and one of these planetary systems, in the star HD~160691 
($\mu$~Arae, e.g. Pepe et al. 2007), contains four planets known so far, 
being the smallest a super-Earth like planet with a minimum 
mass of $\sim 10.5$ Earth masses and its primary planet a 
Jupiter-like planet with $M_p \sin i\sim 1.8$~$M_{\rm Jup}$
(see the extrasolar-planet encyclopaedia\footnote{http://exoplanet.eu}).
We do not really know the nature of this and other super-Earth like 
planets and they may very well be Neptune like ice giants (see e.g.
Barnes et al. 2009).

There is only one star, HD~202206, containing a planet with 
$ M_p \sin i > 10\,M_{\rm Jup}$, which has a primary planet with a 
$M_p \sin i = 17.4$~$M_{\rm Jup}$ at $P_{\rm orb}\sim256$~d, and 
a secondary planet with a $M_p \sin i = 2.44$~$M_{\rm Jup}$ at 
$P_{\rm orb}\sim1383$~d (e.g. Udry et al. 2002; Correia et al. 2005). 
This primary planet may be considered as a \emph{brown dwarf}, 
whose minimum mass may be settled at $\sim 13$~$M_{\rm Jup}$, and/or
entering in the so-called \emph{brown dwarf desert}
(e.g. Halbwachs et al. 2000).

In Fig.~\ref{fallpar} we display the mean abundance ratios 
$\Delta {\rm [X/Fe]_{SUN-STARS}}$ for the 24 stars hosting planets 
according to the orbital period of the primary planet in the 
planetary system of each star. 
We have separated the sample in three ranges: a) 4 stars with primary
planets at $P_{\rm orb} [d] < 25$; b) 10 stars at $150 < P_{\rm
orb}[d] < 650 $; and c) 10 stars at $1000 < P_{\rm orb}[d] < 4300 $.
We note that \emph{Jupiter} has an orbital period of 4333~days 
around the center-of-mass of the solar system.
The distribution of masses of the most massive planet in each 
planetary system as function of orbital periods and metallicities 
are shown in Fig.~\ref{fallmo}. 
There is no case in this sample in which the most massive planet has
orbital period in the gap in between these ranges. Of course,
secondary planets in these systems may lie at orbital period within
these gaps.

Although the linear fits in Fig.~\ref{fallpar} to the element 
abundance ratios seem to change with the orbital period, 
the position of most of these abundance ratios remains the same
and only few elements like Zn, Nd and Zr modify significantly their 
position with respect to other elements and thus changing the
slope of these fits. We suspect that these changes may be 
linked to chemical evolution and abundance scatter, 
and hardly related to the different orbital periods of the 
primary planets. In Table~\ref{tblslope} we give the slopes for
the sub-samples of planet-host stars at different ranges of orbital 
periods. One can appreciate that, due to this large scatter in the 
mean abundance ratios, the derived slopes are very different for 
these different sub-samples. However, the number of stars in these
different sub-samples is so small that we cannot extract any clear 
conclusion from this separation in three orbital period intervals.

\subsection{Metal-rich solar analogs\label{secmr}}

\begin{figure}[!ht]
\centering
\includegraphics[width=5.5cm,angle=90]{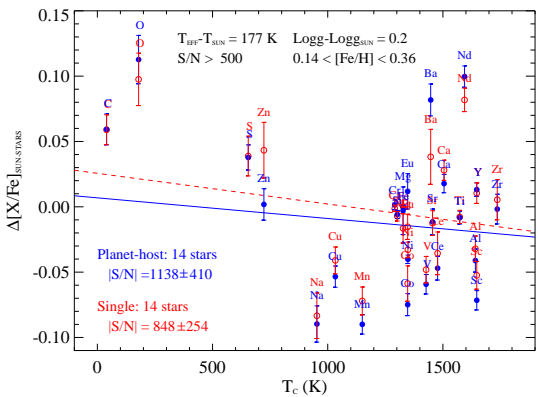}
\caption{Same as in Fig.~\ref{ftwin} for 14 planet host and 
14 ``single'' stars of the sample of metal-rich solar analogs, 
using the \emph{Ganymede} spectrum as solar reference.}  
\label{fmr}
\end{figure}

To evaluate the abundance trends with the same number of planet hosts
and ``single'' stars, we studied a super-solar metallicity sample 
of solar analogs with $T_{\rm eff}-T_{\rm eff,\odot}= \pm
177$~K and $\log g-\log g_\odot=\pm0.2$~dex and $0.15\lesssim
{\rm[Fe/H]} \lesssim 0.4$. This sample contains 14 stars with and 14
without planets. In Fig.~\ref{fmr}, we show the trends 
$\Delta {\rm [X/Fe]_{\rm SUN-STARS}}$ versus $T_C$. The position of
some elements in this plot is certainly affected by chemical evolution
effects due to the high metal content of the sample, in particular ,
Mn and O. This may explain the higher scatter of the points in this
plot with respect to the linear fits. 
However, what is relevant from this plot is that both samples of stars
with and without planets show almost exactly the same abundance
pattern. All the element abundance ratios consistent in both samples
within the error bars, except for the Zn, Mn, Ba, and Nd, which still
are very close in this plot. In addition, the linear fits have almost
exactly the same slope which agrees with the previous statement
(see also Table~\ref{tblslope}). 
It should be mention that among the stars hosting planets, there is a
variety of planetary systems (see Fig.~\ref{fallmo}) with planets 
of different masses at different orbital periods.
In Sect.~\ref{secm} we provide more details on these planetary systems
and discuss the slopes of these trends for each planet-host and
``single'' star, in the context of the presence of terrestial planets.

\subsection{Planetary signatures in the abundance trends?\label{secm}}

Ecuvillon et al. (2006b) measured the slopes of the trends [X/H] 
versus $T_C$ in a sample of 88 planet-host stars and 33 ``single'' 
stars in a large \teff and [Fe/H] ranges, 
$4700 \lesssim T_{\rm eff} {\rm [K]} \lesssim 6400$ and $-0.8\lesssim
{\rm[Fe/H]} \lesssim 0.5$. They did not find any clear difference 
between these two samples. They also searched for
dependence on the planet orbital parameters: mass, orbital period,
eccentricity and orbital separation without any success. 
Ram{\'\i}rez et al. (2009) have derived the slopes in [X/Fe] for two 
$T_C$ ranges above and below 900~K in a sample of solar analogs 
and twins. There appears to be two distinct groups for super-solar 
metallicities, showing positive and negative slopes. Following their
line of reasoning, a null (solar-like) or even negative slope implies 
that a great fraction of refractory elements have been extracted 
from the star-forming cloud to make up dust grains, 
also suggesting planet formation. 
Thus, they tentatively conclude that this bimodal distribution of 
slopes is separating super-solar metallicity stars with and without 
terrestial planets.
Unfortunately, they do not know how many of their stars have planets,
whatever their masses are. One should also note that the core of a
Jupiter-like planet should contain the same amount refractory elements
than roughly three or four terrestial planets. However, Jupiter-like
planets also have a substantial amount of volatiles, whereas
terrestial planets do trap only refractories.

\begin{figure}[!ht]
\centering
\includegraphics[width=5.5cm,angle=90]{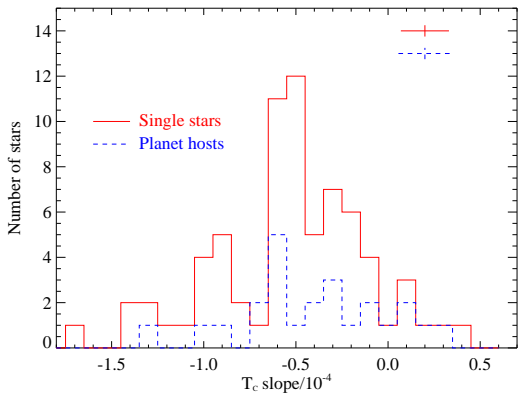}
\caption{Histograms of slopes of linear fits to 71 ``single'' stars 
(solid line) and 24 stars hosting planets (dashed line) 
of the mean element abundance ratios,
 $\Delta {\rm [X/Fe]_{SUN-STARS}}$,
versus the condensation temperature, $T_C$. The mean error bars of
the slopes are shown in upper-right corner.}  
\label{fhisall}
\end{figure}

We have determined the slopes of all solar analogs of our sample with
and without known planets. In Fig.~\ref{fhisall}, we display the
slopes of the trends $\Delta {\rm [X/Fe]_{\rm SUN-STARS}}$ versus the whole
range of $T_C$ for all the stars in our sample. We note that the
slopes in our study have the same sign than in Mel\'endez et al. 
(2009) but opposite than in Ram{\'\i}rez et al. (2009), because the
later used the abundance ratios [X/Fe] instead of 
$\Delta {\rm [X/Fe]_{\rm SUN-STARS}}$. 
Although the number of planet hosts is smaller than ``single'' stars, 
the peak of the distribution of the slopes of these two samples is 
centered around the same position, which corresponds, in fact, 
to a negative slope. We may refer here to the 
Table~\ref{tblslope} where we provide the slope of the mean 
abundance ratios, $\Delta {\rm [X/Fe]_{\rm SUN-STARS}}$, in the whole
range of $T_C$ of the all solar analogs of our sample with and 
without known planets.
These values are slightly higher than the position of the peaks in
Fig.~\ref{fhisall} but the values of the slopes are very similar and
consistent with the error bars for both stars with and without
planets.
This indicates that stars with already detected planets behave 
in similar way, with respect to the chemical abundances, 
as stars without known planets. 
In addition, according to the line of reasoning in Ram{\'\i}rez et al.,
this result also implies that most of
the stars with and without planets in our sample would not have
terrestial planets. However, in two of them, it has been already
detected super-Earth like planets with masses in the range $\sim 7-11$ 
Earth masses. The most massive one is in the planetary system of 
the star HD~160691, with a Jupiter-like planet at 
$P_{\rm orb} \sim 4205$~d and the other one is an isolated planet 
around the star HD~1461 (e.g. Rivera et al. 2010, 
see Fig.~\ref{fallmo}). These two planet-host
stars show clearly a negative slope with --0.57 and --0.27~$\times
10^{-4}$~dex/K which is exactly the opposite to what is expected from
Ram{\'\i}rez et al. (2009). Positive slopes or consistent with zero
within the error bars are found for planet hosts which have either 
only one Jupiter-like planet at $P_{\rm orb} \gtrsim 1200$~d or 
three planetary systems with two planets. Two of them, HD~217107 and
HD~12661 with two Jupiter-like planets each, at very different 
orbital periods (e.g. Wright et al. 2009). 
The other one, HD~47186, contains a Neptune- and a 
Saturn-like planets at $P_{\rm orb} \sim 4$ and $1354$~d,
respectively (Bouchy et al. 2009). 
We may conclude that there is no reason to expect that these stars
hosting relatively massive planets should also contain terrestial 
planets while the other stars with a already detected super-Earth 
like planet should not, and/or that the amount of refractory metals in
the planet hosts depends only on the amount of terrestial planets.
In addition, it seems plausible that many of our targets hosts 
terrestrial planets. This is supported by numerical simulations which
show that terrestrial planets are much more common than giant planets,
and probably 80--90\% of solar-type stars have terrestrial planets 
(e.g. Mordasini et al. 2009a,b). 
This statement agrees with the growing population of low mass planets
found in the HARPS sample of exoplanets (see e.g. 
Udry \& Santos 2007).


\section{Conclusions}

With the aim of investigating possible connection between the
abundance pattern versus the condensation temperature, $T_C$, 
and the presence of terrestial planets, we have selected a sample of 
solar analogs with and without planets from very high signal-to-noise
and very high resolution HARPS, UVES and UES spectra. 
The whole sample contains 71 ``single'' stars and 24 planet hosts 
with $5600 < T_{\rm eff}[{\rm K}] < 5954$,
$ 4.0 <\log (g[{\rm cm~s}^{-2}]) < 4.6$, 
$-0.3 < [{\rm Fe}/{\rm H}] < 0.5$. We perform a detailed chemical
abundance analysis of this sample and investigate possible trends of
mean abundance ratios, $\Delta {\rm [X/Fe]_{\rm SUN-STARS}}$, 
versus $T_C$.
We find that both stars with and without planets show a similar
abundance pattern, except for some elements like Mn which are affected
by chemical evolution effects, due to the higher metal content of the
stars hosting planets.

Only in the sample of HARPS spectra, there are 7 solar twins, 
2 of them hosting planets. We study the mean abundance ratios as a
function of $T_C$ and compare them with the results of Mel\'endez et
al. (2009). Our results are consistent with their results within our
error bars, which has been demonstrated through two Monte Carlo
simulations. Nevertheless, the scatter around the linear fits in 
our data is larger, which may be related to the solar reference 
HARPS spectrum used in this analysis.

We also select a metal-rich sample of solar analogs in the range
$[{\rm Fe}/{\rm H}] = 0.25 \pm 0.11$~dex. This sample has 14 stars
hosting planets and 14 ``single'' stars. In this case, the abundance
pattern and the slope of the linear fit to both samples are 
almost equal, which may indicate that the abundance pattern found by 
Mel\'endez et al. (2009) do not have nothing to do with the presence
of planets.

Finally, we derive the slopes of the trend 
$\Delta {\rm [X/Fe]_{\rm SUN-STARS}}$ versus $T_C$ for each star 
of the whole sample and
compare the distribution of the slopes found for stars with and 
without planets. Although the number of planet hosts is significantly
smaller than that of ``single'' stars, the peaks of the 
distributions are placed at the same position, at about $-0.5 \times
10^{-4}$~dex/K, which is exactly the opposite sign than that required by
the presence of terrestial planets, following the line of reasoning 
in Ram{\'\i}rez et al. (2009). Thus, according to Ram{\'\i}rez et al. 
(2009), most of our planet-host stars would not contain terrestial
planets, a statement that a priori does not appear to be expected. 
  
\acknowledgments

J.I.G.H. acknowledges financial support from the Spanish Ministry 
project MICINN AYA2008-00695 and also from the Spanish Ministry of
Science and Innovation (MICINN) under the 2009 Juan de la Cierva
Programme. 
N.C.S., S.S. and V.N. would like to thank the support by the 
European Research Council/European Community under the FP7 through a 
Starting Grant, as well from Funda\c{c}\~ao para a Ci\^encia e a
Tecnologia (FCT), Portugal, through a Ci\^encia\,2007 
contract funded by FCT/MCTES (Portugal) and POPH/FSE (EC), 
and in the form of grant reference PTDC/CTE-AST/098528/2008
from FCT/MCTES.
E.D.M, J.I.G.H. and G.I. would like to thank financial
support from the Spanish Ministry project MICINN AYA2008-04874.
S.S. acknowledges the support of the FCT grant: SFRH/BPD/47611/2008. 
This work has also made use of
the IRAF facility, and the Encyclopaedia of extrasolar planets.

\section{Online Material}

\clearpage

\begin{table*}
\centering
\scriptsize
\caption{Abundance ratios [X/Fe] of solar analogs with planets\label{tabpla1}}
\begin{tabular}{lrrrrrr}
\noalign{\smallskip}
\noalign{\smallskip}
\noalign{\smallskip}
\hline
\hline
\noalign{\medskip} 
HD       &     [C/Fe]      &     [O/Fe]      &     [S/Fe]      &     [Na/Fe]      &     [Mg/Fe]      &     [Al/Fe]      \\
\noalign{\medskip} 
\hline
\hline
\noalign{\medskip} 
  102117 & $-0.046\pm0.007$ & $-0.101\pm0.070$ & $ 0.004\pm0.064$ & $ 0.054\pm0.035$ & $ 0.024\pm0.078$ & $ 0.074\pm0.035$ \\
  114729 & $ 0.075\pm0.028$ & $ 0.206\pm0.070$ & $ 0.066\pm0.028$ & $-0.009\pm0.049$ & $ 0.171\pm0.106$ & $ 0.081\pm0.035$ \\
  134987 & $ 0.032\pm0.028$ & $-0.048\pm0.070$ & $ 0.017\pm0.007$ & $ 0.137\pm0.021$ & $ 0.017\pm0.078$ & $ 0.057\pm0.035$ \\
    1461 & $-0.037\pm0.028$ & $-0.127\pm0.070$ & $-0.047\pm0.070$ & $ 0.058\pm0.049$ & $-0.002\pm0.049$ & $ 0.033\pm0.028$ \\
  160691 & $-0.047\pm0.007$ & $-0.052\pm0.070$ & $-0.037\pm0.021$ & $ 0.038\pm0.113$ & $ 0.038\pm0.042$ & $ 0.043\pm0.035$ \\
   16417 & $-0.080\pm0.028$ & $-0.010\pm0.070$ & $-0.090\pm0.042$ & $-0.025\pm0.106$ & $ 0.045\pm0.035$ & $ 0.015\pm0.035$ \\
  196050 & $-0.021\pm0.028$ & $ 0.159\pm0.070$ & $-0.031\pm0.042$ & $ 0.109\pm0.071$ & $ 0.059\pm0.014$ & $ 0.059\pm0.028$ \\
\hline
\end{tabular}
\end{table*}

\clearpage

\begin{table*}
\centering
\scriptsize
\caption{Abundance ratios [X/Fe] of solar analogs with planets\label{tabpla2}}
\begin{tabular}{lrrrrrr}
\noalign{\smallskip}
\noalign{\smallskip}
\noalign{\smallskip}
\hline
\hline
\noalign{\medskip} 
HD       &     [Si/Fe]      &     [Ca/Fe]      &     [Sc/Fe]      &     [Ti/Fe]      &     [V/Fe]      &     [Cr/Fe]      \\
\noalign{\medskip} 
\hline
\hline
\noalign{\medskip} 
  102117 & $ 0.018\pm0.020$ & $ 0.007\pm0.062$ & $ 0.101\pm0.061$ & $ 0.021\pm0.036$ & $ 0.059\pm0.021$ & $-0.012\pm0.024$ \\
  114729 & $ 0.074\pm0.017$ & $ 0.118\pm0.032$ & $ 0.120\pm0.026$ & $ 0.104\pm0.024$ & $ 0.012\pm0.038$ & $ 0.009\pm0.023$ \\
  134987 & $ 0.029\pm0.027$ & $-0.025\pm0.032$ & $ 0.087\pm0.055$ & $-0.001\pm0.033$ & $ 0.050\pm0.015$ & $ 0.002\pm0.022$ \\
    1461 & $-0.002\pm0.021$ & $-0.021\pm0.055$ & $ 0.083\pm0.029$ & $ 0.004\pm0.015$ & $ 0.048\pm0.026$ & $-0.000\pm0.009$ \\
  160691 & $ 0.008\pm0.016$ & $-0.051\pm0.055$ & $ 0.091\pm0.052$ & $ 0.010\pm0.030$ & $ 0.046\pm0.021$ & $ 0.007\pm0.017$ \\
   16417 & $-0.005\pm0.012$ & $ 0.038\pm0.046$ & $ 0.035\pm0.029$ & $ 0.000\pm0.019$ & $-0.005\pm0.023$ & $ 0.002\pm0.013$ \\
  196050 & $ 0.018\pm0.020$ & $-0.017\pm0.028$ & $ 0.123\pm0.049$ & $ 0.011\pm0.016$ & $ 0.044\pm0.019$ & $ 0.001\pm0.021$ \\
\hline
\end{tabular}
\end{table*}

\clearpage

\begin{table*}
\centering
\scriptsize
\caption{Abundance ratios [X/Fe] of solar analogs with planets\label{tabpla3}}
\begin{tabular}{lrrrrrr}
\noalign{\smallskip}
\noalign{\smallskip}
\noalign{\smallskip}
\hline
\hline
\noalign{\medskip} 
HD       &     [Mn/Fe]      &     [Co/Fe]      &     [Ni/Fe]      &     [Cu/Fe]      &     [Zn/Fe]      &     [Sr/Fe]      \\
\noalign{\medskip} 
\hline
\hline
\noalign{\medskip} 
  102117 & $ 0.079\pm0.043$ & $ 0.099\pm0.042$ & $ 0.040\pm0.025$ & $ 0.049\pm0.070$ & $ 0.024\pm0.078$ & $-0.001\pm0.070$ \\
  114729 & $-0.172\pm0.061$ & $-0.006\pm0.026$ & $-0.031\pm0.019$ & $ 0.056\pm0.070$ & $ 0.066\pm0.127$ & $-0.104\pm0.070$ \\
  134987 & $ 0.062\pm0.068$ & $ 0.109\pm0.012$ & $ 0.055\pm0.017$ & $ 0.092\pm0.070$ & $ 0.052\pm0.057$ & $-0.018\pm0.070$ \\
    1461 & $ 0.093\pm0.016$ & $ 0.060\pm0.020$ & $ 0.038\pm0.016$ & $ 0.053\pm0.070$ & $ 0.028\pm0.106$ & $ 0.003\pm0.070$ \\
  160691 & $ 0.098\pm0.029$ & $ 0.078\pm0.018$ & $ 0.031\pm0.027$ & $ 0.078\pm0.070$ & $ 0.058\pm0.057$ & $-0.022\pm0.070$ \\
   16417 & $ 0.010\pm0.020$ & $ 0.015\pm0.017$ & $ 0.004\pm0.016$ & $ 0.020\pm0.070$ & $-0.015\pm0.064$ & $-0.080\pm0.070$ \\
  196050 & $ 0.069\pm0.016$ & $ 0.074\pm0.021$ & $ 0.043\pm0.014$ & $ 0.069\pm0.070$ & $-0.016\pm0.021$ & $-0.021\pm0.070$ \\
\hline
\end{tabular}
\end{table*}

\clearpage

\begin{table*}
\centering
\scriptsize
\caption{Abundance ratios [X/Fe] of solar analogs with planets\label{tabpla4}}
\begin{tabular}{lrrrrrr}
\noalign{\smallskip}
\noalign{\smallskip}
\noalign{\smallskip}
\hline
\hline
\noalign{\medskip} 
HD       &     [Y/Fe]      &     [Zr/Fe]      &     [Ba/Fe]      &     [Ce/Fe]      &     [Nd/Fe]      &     [Eu/Fe]      \\
\noalign{\medskip} 
\hline
\hline
\noalign{\medskip} 
  102117 & $-0.015\pm0.049$ & $ 0.049\pm0.070$ & $-0.116\pm0.134$ & $ 0.104\pm0.078$ & $-0.051\pm0.070$ & $ 0.099\pm0.070$ \\
  114729 & $-0.044\pm0.035$ & $ 0.016\pm0.070$ & $ 0.066\pm0.099$ &        --        & $-0.025\pm0.070$ & $ 0.155\pm0.070$ \\
  134987 & $-0.038\pm0.061$ & $ 0.022\pm0.070$ & $-0.098\pm0.070$ & $ 0.057\pm0.078$ & $-0.138\pm0.070$ & $ 0.012\pm0.070$ \\
    1461 & $-0.017\pm0.052$ & $-0.077\pm0.070$ & $-0.057\pm0.070$ & $ 0.028\pm0.078$ & $-0.087\pm0.070$ & $ 0.003\pm0.070$ \\
  160691 & $-0.032\pm0.069$ & $-0.002\pm0.070$ & $-0.147\pm0.148$ & $ 0.053\pm0.035$ & $-0.102\pm0.070$ & $ 0.018\pm0.070$ \\
   16417 & $-0.023\pm0.021$ & $-0.060\pm0.070$ & $-0.040\pm0.028$ & $-0.010\pm0.070$ & $-0.100\pm0.070$ & $ 0.010\pm0.070$ \\
  196050 & $-0.014\pm0.021$ & $-0.011\pm0.070$ & $-0.066\pm0.007$ & $-0.001\pm0.042$ & $-0.121\pm0.070$ & $-0.001\pm0.070$ \\
\hline
\end{tabular}
\end{table*}

\clearpage

\begin{table*}
\centering
\scriptsize
\caption{Abundance ratios [X/Fe] of solar analogs without known planets\label{tabsin1}}
\begin{tabular}{lrrrrrr}
\noalign{\smallskip}
\noalign{\smallskip}
\noalign{\smallskip}
\hline
\hline
\noalign{\medskip} 
HD       &     [C/Fe]      &     [O/Fe]      &     [S/Fe]      &     [Na/Fe]      &     [Mg/Fe]      &     [Al/Fe]      \\
\noalign{\medskip} 
\hline
\hline
\noalign{\medskip} 
   10180 & $ 0.011\pm0.035$ & $-0.014\pm0.070$ & $ 0.011\pm0.021$ & $ 0.071\pm0.049$ & $-0.019\pm0.078$ & $ 0.006\pm0.070$ \\
  102365 & $ 0.072\pm0.113$ & $ 0.252\pm0.070$ & $ 0.037\pm0.134$ & $ 0.012\pm0.057$ & $ 0.092\pm0.014$ & $ 0.122\pm0.014$ \\
  104982 & $ 0.003\pm0.021$ & $-0.022\pm0.070$ & $-0.147\pm0.134$ & $ 0.023\pm0.049$ & $ 0.043\pm0.021$ & $ 0.038\pm0.042$ \\
  106116 & $-0.053\pm0.021$ & $-0.018\pm0.070$ & $-0.048\pm0.014$ & $ 0.012\pm0.071$ & $-0.013\pm0.064$ & $ 0.027\pm0.035$ \\
  108309 & $ 0.023\pm0.042$ & $ 0.003\pm0.070$ & $ 0.023\pm0.071$ & $ 0.023\pm0.085$ & $ 0.003\pm0.071$ & $ 0.023\pm0.028$ \\
  109409 & $-0.066\pm0.021$ & $-0.081\pm0.070$ & $-0.016\pm0.021$ & $ 0.154\pm0.049$ & $-0.021\pm0.127$ & $ 0.029\pm0.028$ \\
  111031 & $-0.045\pm0.021$ & $-0.009\pm0.070$ & $-0.060\pm0.014$ & $ 0.090\pm0.014$ & $-0.030\pm0.099$ & $-0.030\pm0.071$ \\
\hline
\end{tabular}
\end{table*}

\clearpage

\begin{table*}
\centering
\scriptsize
\caption{Continued.} 
\begin{tabular}{lrrrrrr}
\noalign{\smallskip}
\noalign{\smallskip}
\noalign{\smallskip}
\hline
\hline
\noalign{\medskip} 
HD       &     [C/Fe]      &     [O/Fe]      &     [S/Fe]      &     [Na/Fe]      &     [Mg/Fe]      &     [Al/Fe]      \\
\noalign{\medskip} 
\hline
\hline
\noalign{\medskip} 
  210918 & $ 0.050\pm0.035$ & $ 0.175\pm0.070$ & $-0.035\pm0.028$ & $-0.035\pm0.028$ & $ 0.020\pm0.078$ & $ 0.070\pm0.007$ \\
  212708 & $-0.030\pm0.007$ & $-0.075\pm0.070$ & $-0.005\pm0.014$ & $ 0.095\pm0.028$ & $-0.030\pm0.120$ & $ 0.055\pm0.028$ \\
  221146 & $ 0.002\pm0.007$ & $-0.103\pm0.070$ & $-0.108\pm0.120$ & $ 0.057\pm0.057$ & $ 0.022\pm0.035$ & $ 0.007\pm0.028$ \\
  222595 & $-0.060\pm0.042$ & $-0.050\pm0.070$ & $-0.030\pm0.028$ & $-0.065\pm0.049$ & $-0.035\pm0.035$ & $-0.025\pm0.007$ \\
  222669 & $-0.058\pm0.014$ & $-0.068\pm0.070$ & $-0.087\pm0.014$ & $-0.003\pm0.035$ & $ 0.027\pm0.078$ & $-0.072\pm0.007$ \\
  223171 & $-0.008\pm0.035$ & $ 0.057\pm0.070$ & $-0.028\pm0.021$ & $ 0.002\pm0.106$ & $ 0.032\pm0.049$ & $ 0.017\pm0.028$ \\
   27063 & $-0.084\pm0.007$ & $-0.069\pm0.070$ & $-0.074\pm0.007$ & $-0.084\pm0.092$ & $-0.014\pm0.007$ & $-0.049\pm0.014$ \\
\hline
\end{tabular}
\end{table*}

\clearpage

\begin{table*}
\centering
\scriptsize
\caption{Abundance ratios [X/Fe] of solar analogs without known planets\label{tabsin2}}
\begin{tabular}{lrrrrrr}
\noalign{\smallskip}
\noalign{\smallskip}
\noalign{\smallskip}
\hline
\hline
\noalign{\medskip} 
HD       &     [Si/Fe]      &     [Ca/Fe]      &     [Sc/Fe]      &     [Ti/Fe]      &     [V/Fe]      &     [Cr/Fe]      \\
\noalign{\medskip} 
\hline
\hline
\noalign{\medskip} 
   10180 & $ 0.022\pm0.010$ & $ 0.004\pm0.018$ & $ 0.056\pm0.040$ & $-0.003\pm0.017$ & $-0.009\pm0.029$ & $ 0.004\pm0.012$ \\
  102365 & $ 0.104\pm0.038$ & $ 0.051\pm0.034$ & $ 0.112\pm0.085$ & $ 0.130\pm0.028$ & $ 0.046\pm0.032$ & $-0.004\pm0.036$ \\
  104982 & $ 0.021\pm0.013$ & $ 0.036\pm0.010$ & $ 0.015\pm0.029$ & $ 0.038\pm0.020$ & $ 0.020\pm0.015$ & $ 0.002\pm0.011$ \\
  106116 & $-0.004\pm0.026$ & $-0.004\pm0.032$ & $ 0.028\pm0.027$ & $ 0.012\pm0.019$ & $ 0.026\pm0.027$ & $-0.000\pm0.007$ \\
  108309 & $-0.001\pm0.017$ & $-0.005\pm0.007$ & $ 0.038\pm0.021$ & $-0.001\pm0.021$ & $-0.004\pm0.019$ & $ 0.004\pm0.010$ \\
  109409 & $-0.008\pm0.034$ & $-0.058\pm0.021$ & $ 0.071\pm0.036$ & $-0.010\pm0.039$ & $ 0.063\pm0.021$ & $-0.016\pm0.017$ \\
  111031 & $-0.005\pm0.022$ & $-0.061\pm0.042$ & $ 0.105\pm0.031$ & $ 0.004\pm0.019$ & $ 0.059\pm0.016$ & $-0.008\pm0.020$ \\
\hline
\end{tabular}
\end{table*}

\clearpage

\begin{table*}
\centering
\scriptsize
\caption{Continued.} 
\begin{tabular}{lrrrrrr}
\noalign{\smallskip}
\noalign{\smallskip}
\noalign{\smallskip}
\hline
\hline
\noalign{\medskip} 
HD       &     [C/Fe]      &     [O/Fe]      &     [S/Fe]      &     [Na/Fe]      &     [Mg/Fe]      &     [Al/Fe]      \\
\noalign{\medskip} 
\hline
\hline
\noalign{\medskip} 
  210918 & $ 0.027\pm0.013$ & $ 0.058\pm0.008$ & $ 0.067\pm0.022$ & $ 0.070\pm0.020$ & $-0.004\pm0.021$ & $ 0.000\pm0.012$ \\
  212708 & $ 0.011\pm0.025$ & $-0.057\pm0.030$ & $ 0.095\pm0.028$ & $ 0.009\pm0.030$ & $ 0.076\pm0.037$ & $ 0.001\pm0.020$ \\
  221146 & $ 0.025\pm0.016$ & $ 0.034\pm0.027$ & $ 0.027\pm0.014$ & $-0.018\pm0.030$ & $-0.030\pm0.046$ & $-0.002\pm0.010$ \\
  222595 & $-0.004\pm0.016$ & $ 0.022\pm0.010$ & $-0.014\pm0.025$ & $ 0.010\pm0.016$ & $ 0.002\pm0.015$ & $ 0.011\pm0.017$ \\
  222669 & $-0.013\pm0.012$ & $ 0.014\pm0.015$ & $-0.030\pm0.033$ & $-0.027\pm0.024$ & $-0.041\pm0.019$ & $ 0.008\pm0.014$ \\
  223171 & $ 0.000\pm0.013$ & $ 0.013\pm0.043$ & $ 0.052\pm0.029$ & $-0.005\pm0.018$ & $-0.001\pm0.015$ & $-0.000\pm0.019$ \\
   27063 & $-0.018\pm0.014$ & $ 0.025\pm0.019$ & $-0.032\pm0.021$ & $ 0.000\pm0.020$ & $-0.009\pm0.017$ & $ 0.011\pm0.013$ \\
\hline
\end{tabular}
\end{table*}

\clearpage

\begin{table*}
\centering
\scriptsize
\caption{Abundance ratios [X/Fe] of solar analogs without known planets\label{tabsin3}}
\begin{tabular}{lrrrrrr}
\noalign{\smallskip}
\noalign{\smallskip}
\noalign{\smallskip}
\hline
\hline
\noalign{\medskip} 
HD       &     [Mn/Fe]      &     [Co/Fe]      &     [Ni/Fe]      &     [Cu/Fe]      &     [Zn/Fe]      &     [Sr/Fe]      \\
\noalign{\medskip} 
\hline
\hline
\noalign{\medskip} 
   10180 & $ 0.006\pm0.008$ & $ 0.026\pm0.013$ & $ 0.021\pm0.014$ & $ 0.036\pm0.070$ & $-0.034\pm0.028$ & $-0.044\pm0.070$ \\
  102365 & $-0.171\pm0.043$ & $ 0.026\pm0.018$ & $-0.008\pm0.042$ & $-0.008\pm0.070$ & $ 0.022\pm0.085$ & $-0.128\pm0.070$ \\
  104982 & $-0.107\pm0.037$ & $ 0.003\pm0.019$ & $-0.020\pm0.013$ & $ 0.018\pm0.070$ & $-0.002\pm0.014$ & $-0.092\pm0.070$ \\
  106116 & $ 0.065\pm0.030$ & $ 0.044\pm0.026$ & $ 0.020\pm0.013$ & $ 0.022\pm0.070$ & $ 0.012\pm0.085$ & $ 0.022\pm0.070$ \\
  108309 & $ 0.038\pm0.013$ & $ 0.029\pm0.024$ & $-0.005\pm0.015$ & $ 0.033\pm0.070$ & $ 0.053\pm0.099$ & $-0.117\pm0.070$ \\
  109409 & $ 0.126\pm0.026$ & $ 0.092\pm0.028$ & $ 0.061\pm0.018$ & $ 0.069\pm0.070$ & $ 0.064\pm0.148$ & $-0.041\pm0.070$ \\
  111031 & $ 0.098\pm0.030$ & $ 0.085\pm0.026$ & $ 0.045\pm0.022$ & $ 0.060\pm0.070$ & $ 0.051\pm0.099$ & $ 0.010\pm0.070$ \\
\hline
\end{tabular}
\end{table*}

\clearpage

\begin{table*}
\centering
\scriptsize
\caption{Continued.} 
\begin{tabular}{lrrrrrr}
\noalign{\smallskip}
\noalign{\smallskip}
\noalign{\smallskip}
\hline
\hline
\noalign{\medskip} 
HD       &     [C/Fe]      &     [O/Fe]      &     [S/Fe]      &     [Na/Fe]      &     [Mg/Fe]      &     [Al/Fe]      \\
\noalign{\medskip} 
\hline
\hline
\noalign{\medskip} 
  210918 & $-0.125\pm0.048$ & $ 0.002\pm0.018$ & $-0.035\pm0.012$ & $ 0.025\pm0.070$ & $ 0.050\pm0.035$ & $-0.085\pm0.070$ \\
  212708 & $ 0.077\pm0.048$ & $ 0.106\pm0.031$ & $ 0.051\pm0.022$ & $ 0.065\pm0.070$ & $ 0.055\pm0.141$ & $ 0.035\pm0.070$ \\
  221146 & $ 0.003\pm0.045$ & $ 0.019\pm0.007$ & $ 0.007\pm0.020$ & $ 0.057\pm0.070$ & $ 0.037\pm0.085$ & $ 0.147\pm0.070$ \\
  222595 & $ 0.014\pm0.034$ & $ 0.008\pm0.028$ & $-0.008\pm0.013$ & $-0.020\pm0.070$ & $-0.005\pm0.049$ & $ 0.100\pm0.070$ \\
  222669 & $-0.015\pm0.015$ & $-0.043\pm0.008$ & $-0.020\pm0.012$ & $-0.038\pm0.070$ & $-0.047\pm0.028$ & $ 0.052\pm0.070$ \\
  223171 & $-0.000\pm0.022$ & $ 0.029\pm0.018$ & $ 0.004\pm0.014$ & $ 0.027\pm0.070$ & $-0.033\pm0.070$ & $-0.083\pm0.070$ \\
   27063 & $-0.004\pm0.013$ & $-0.026\pm0.018$ & $-0.026\pm0.012$ & $-0.039\pm0.070$ & $-0.054\pm0.078$ & $ 0.111\pm0.070$ \\
\hline
\end{tabular}
\end{table*}

\clearpage

\begin{table*}
\centering
\scriptsize
\caption{Abundance ratios [X/Fe] of solar analogs without known planets\label{tabsin4}}
\begin{tabular}{lrrrrrr}
\noalign{\smallskip}
\noalign{\smallskip}
\noalign{\smallskip}
\hline
\hline
\noalign{\medskip} 
HD       &     [Y/Fe]      &     [Zr/Fe]      &     [Ba/Fe]      &     [Ce/Fe]      &     [Nd/Fe]      &     [Eu/Fe]      \\
\noalign{\medskip} 
\hline
\hline
\noalign{\medskip} 
   10180 & $ 0.016\pm0.026$ & $ 0.016\pm0.070$ & $-0.024\pm0.014$ & $-0.009\pm0.021$ & $-0.134\pm0.070$ & $-0.024\pm0.070$ \\
  102365 &        --        & $ 0.152\pm0.070$ &        --        & $-0.028\pm0.071$ & $ 0.022\pm0.070$ & $ 0.102\pm0.070$ \\
  104982 & $-0.026\pm0.025$ & $ 0.008\pm0.070$ & $ 0.113\pm0.120$ & $-0.022\pm0.085$ & $-0.042\pm0.070$ & $ 0.068\pm0.070$ \\
  106116 & $-0.031\pm0.049$ & $-0.018\pm0.070$ & $-0.023\pm0.021$ & $ 0.052\pm0.071$ & $-0.088\pm0.070$ & $ 0.002\pm0.070$ \\
  108309 & $-0.047\pm0.020$ & $-0.097\pm0.070$ & $-0.027\pm0.070$ & $-0.002\pm0.007$ & $-0.147\pm0.070$ & $-0.007\pm0.070$ \\
  109409 & $-0.048\pm0.055$ & $-0.031\pm0.070$ & $-0.106\pm0.049$ & $-0.021\pm0.042$ & $-0.141\pm0.070$ & $ 0.029\pm0.070$ \\
  111031 & $-0.006\pm0.057$ & $-0.100\pm0.070$ & $-0.035\pm0.035$ & $ 0.051\pm0.085$ & $-0.100\pm0.070$ & $ 0.040\pm0.070$ \\
\hline
\end{tabular}
\end{table*}

\clearpage

\begin{table*}
\centering
\scriptsize
\caption{Continued.} 
\begin{tabular}{lrrrrrr}
\noalign{\smallskip}
\noalign{\smallskip}
\noalign{\smallskip}
\hline
\hline
\noalign{\medskip} 
HD       &     [C/Fe]      &     [O/Fe]      &     [S/Fe]      &     [Na/Fe]      &     [Mg/Fe]      &     [Al/Fe]      \\
\noalign{\medskip} 
\hline
\hline
\noalign{\medskip} 
  210918 & $-0.025\pm0.026$ & $ 0.055\pm0.070$ & $ 0.040\pm0.007$ & $ 0.045\pm0.028$ & $-0.005\pm0.070$ & $ 0.115\pm0.070$ \\
  212708 & $-0.035\pm0.062$ & $ 0.055\pm0.070$ & $-0.165\pm0.113$ & $ 0.090\pm0.078$ & $-0.075\pm0.070$ & $ 0.035\pm0.070$ \\
  221146 & $ 0.120\pm0.038$ & $ 0.047\pm0.070$ & $ 0.037\pm0.014$ & $-0.033\pm0.014$ & $-0.093\pm0.070$ & $ 0.037\pm0.070$ \\
  222595 & $ 0.060\pm0.010$ & $ 0.070\pm0.070$ & $ 0.080\pm0.070$ & $ 0.075\pm0.064$ & $-0.080\pm0.070$ & $-0.040\pm0.070$ \\
  222669 & $ 0.079\pm0.021$ & $-0.008\pm0.070$ & $ 0.117\pm0.049$ & $ 0.037\pm0.007$ & $-0.108\pm0.070$ & $ 0.002\pm0.070$ \\
  223171 & $-0.023\pm0.020$ & $-0.053\pm0.070$ & $ 0.022\pm0.021$ & $-0.013\pm0.028$ & $-0.123\pm0.070$ & $-0.013\pm0.070$ \\
   27063 & $ 0.091\pm0.017$ & $ 0.091\pm0.070$ & $ 0.141\pm0.070$ & $ 0.086\pm0.021$ & $-0.039\pm0.070$ & $-0.019\pm0.070$ \\
\hline
\end{tabular}
\end{table*}

\clearpage

\end{document}